\documentclass[10pt,twocolumn,letterpaper,pagenumbers]{article}

\usepackage[pagenumbers]{iccv}

\usepackage{amssymb} 
\usepackage{pifont} 
\newcommand{\cmark}{\ding{51}}%
\newcommand{\xmark}{\ding{55}}%
\usepackage{wrapfig}
\usepackage{graphicx} %
\usepackage{sidecap, caption}
\usepackage{tikz} 
\usepackage{xcolor} 
\usepackage{multirow}

\usepackage{tocloft}

\newcommand{\ourmethod}{{{GarmentCrafter}}\xspace}
\newcommand\mypara[1]{\vspace{1mm}\noindent\textbf{#1}}

\definecolor{myblue}{rgb}{0.21,0.49,0.74}
\usepackage[pagebackref,breaklinks,colorlinks,allcolors=myblue]{hyperref}

\title{\ourmethod: Progressive Novel View Synthesis for Single-View \\ 3D Garment Reconstruction and Editing}

\author{
Yuanhao Wang$^{1}$ \quad
Cheng Zhang$^{2}$ \quad
Gonçalo Frazão$^{1}$ \quad
Jinlong Yang$^{3}$ \quad
Alexandru-Eugen Ichim$^{3}$ \\  [2pt]
Thabo Beeler$^{3}$ \quad
Fernando De la Torre$^{1}$ \\ [6pt]
$^{1}$ {Carnegie Mellon University} \quad
$^{2}$ {Texas A\&M University} \quad
$^{3}$ {Google AR} \\ [6pt]
\href{https:/humansensinglab.github.io/garment-crafter/}{humansensinglab.github.io/garment-crafter}
}

\begin{document}
\twocolumn[{%
\renewcommand\twocolumn[1][]{#1}%
\maketitle
\begin{center}
    \vspace{-3.5mm}
    \includegraphics[width=0.995\textwidth]{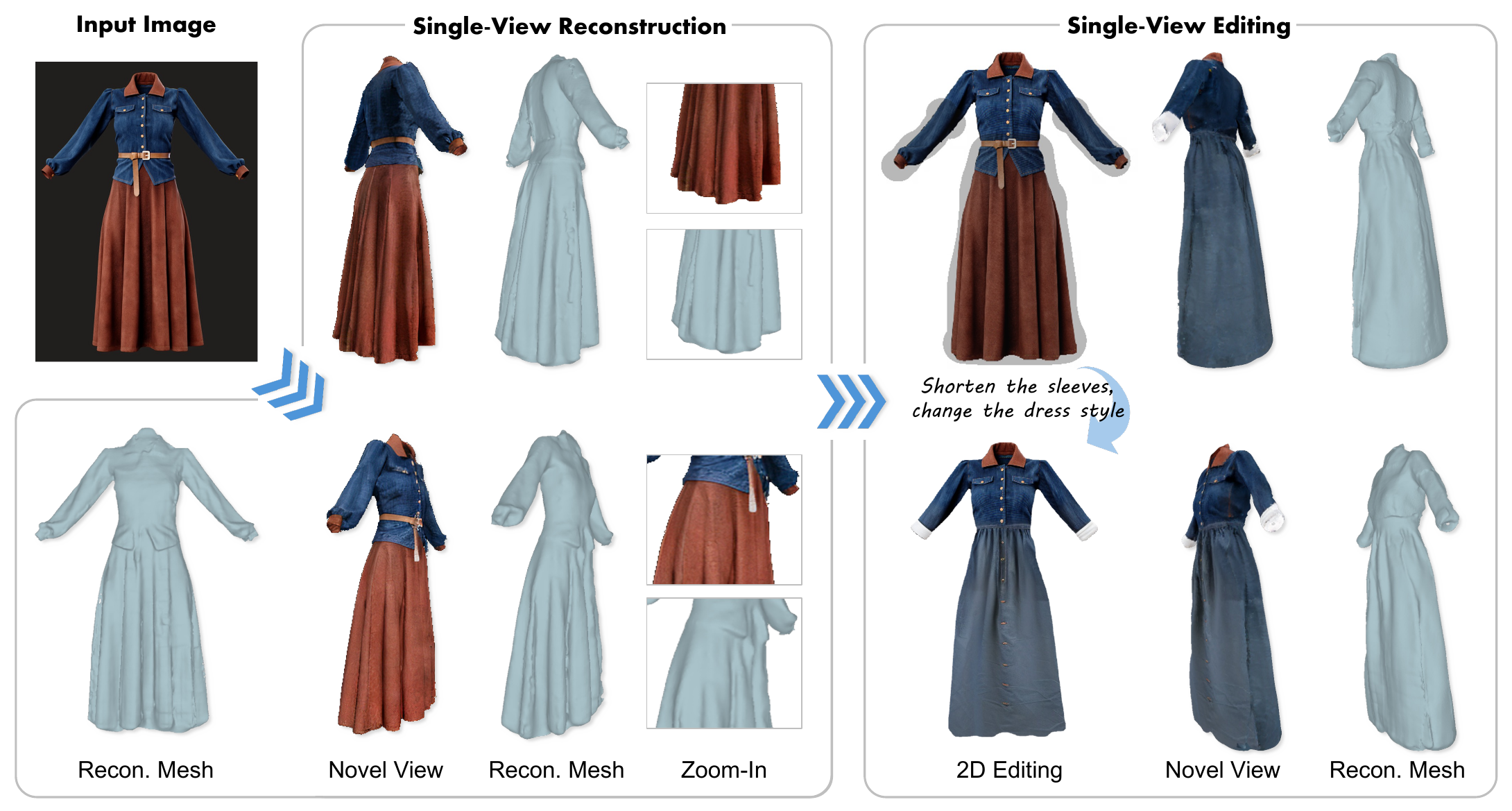}
    \vspace{-3mm}
    \captionof{figure}{From a real-world clothing image, \ourmethod synthesizes high-quality novel views, enabling the reconstruction of garment meshes with accurate geometry and rich detail. Additionally, users can easily apply 2D edits (e.g., modifying parts or surface details) using off-the-shelf tools on a single image, and \ourmethod seamlessly applies these edits across the 3D model with multi-view consistency.}
    \label{fig:teaser}
    \vspace{2.5mm}
\end{center}
}]

\begin{abstract}
We introduce \ourmethod, a new approach that enables non-professional users to create and modify 3D garments from a single-view image. While recent advances in image generation have facilitated 2D garment design, creating and editing 3D garments remains challenging for non-professional users. Existing methods for single-view 3D reconstruction often rely on pre-trained generative models to synthesize novel views conditioning on the reference image and camera pose, yet they lack cross-view consistency, failing to capture the internal relationships across different views. In this paper, we tackle this challenge through progressive depth prediction and image warping to approximate novel views. Subsequently, we train a multi-view diffusion model to complete occluded and unknown clothing regions, informed by the evolving camera pose. By jointly inferring RGB and depth, \ourmethod enforces inter-view coherence and reconstructs precise geometries and fine details. Extensive experiments demonstrate that our method achieves superior visual fidelity and inter-view coherence compared to state-of-the-art single-view 3D garment reconstruction methods. 
\end{abstract}

\section{Introduction}
\label{sec:intro}

Professional fashion designers use sophisticated software to create and edit garments in 3D, crafting highly detailed virtual apparels~\cite{Clo3D,TUKA3D,browzwear,Style3D}. However, as digital garments become integral to virtual environments and personalized digital experiences~\cite{hu2024gaussianavatar, casado2022pergamo, xiu2024puzzleavatar,guo2023vid2avatar,rong2024gaussian}, there is a growing demand for intuitive tools that allow non-professional users to design and interact with 3D garments. For broader accessibility, such tools should allow users to work with 3D garments with minimal input, ideally from just a single image. This raises a key question: \emph{How can we create and edit 3D garments with simple manipulations in an image?}

Recent advancements in image generation models~\cite{rombach2022high,ramesh2021zero,saharia2022photorealistic,wang2024towards} and image editing techniques~\cite{zhou2010parametric,brooks2023instructpix2pix,pan2023drag,zhu2024m,wang2024texfit,raj2018swapnet} have enabled high-quality garment design in 2D. Yet, achieving the same level of control and realism for 3D garments remains challenging for common users. Currently, state-of-the-art methods on single-view 3D garments rely either on 1) deforming, matching, and registration with the human body prior~\cite{loper2023smpl} and/or predefined garment templates~\cite{liu2023towards, sarafianos2024garment3dgen,corona2021smplicit,li2024garment,gao2023cloth2tex,majithia2022robust,bhatnagar2019multi}, or 2) novel view synthesis techniques~\cite{watson2022novel,liu2023zero} that use pre-trained 2D diffusion models conditioned on a reference image and target pose. However, they often fall short in capturing accurate, realistic geometry and appearance.

Two characteristics of garments pose challenges. First, garments exhibit diverse shapes, complex geometries, and rich textures, making template-based methods limited in their ability to generalize across clothing styles. Most existing methods prioritize either geometry~\cite{luo2024garverselod,corona2021smplicit} or texture~\cite{zhang2024sa,richardson2023texture}, rarely balancing both~\cite{sarafianos2024garment3dgen,gao2023cloth2tex,majithia2022robust}. Second, the fine details in garments demand stronger multi-view consistency. Existing novel view synthesis methods~\cite{xu2024instantmesh,liu2023syncdreamer}, conditioned on a reference image and target pose, often neglect critical semantic connections across different views.

How can we ensure that a pixel in one view corresponds to a point visible in another, with consistent appearance? In this paper, we propose a different approach, \emph{progressive novel view synthesis}, to enhance cross-view coherence. Our method begins by estimating the depth of the input image and warping projected points to approximate unseen views. We then apply a multi-view diffusion model to complete missing and occluded regions based on the evolving camera pose. Furthermore, we incorporate a monocular depth estimation model to generate depth maps that remain consistent with the warped depths. Unlike existing novel view synthesis, our key insight is to use the depth-based warped image as an additional condition to guide cross-view alignment. By progressively synthesizing views and depths along a predefined camera trajectory, our method gradually refines the geometry and texture of the garment across viewpoints.

We name our method \emph{\ourmethod}, a novel solution for 3D garment creation and editing while users just need to operate on a single-view image, as shown in Figure~\ref{fig:teaser}. Specifically, \ourmethod not only generates high-quality 3D garments but also extends garment editing from 2D to 3D. 
Thanks to our progressive novel view synthesis, users can make local edits (e.g., editing surface details) or perform part-based manipulations (e.g., modifying garment parts) directly on a single-view image, with precise effects reflected in 3D space --- capabilities that are absent in the existing methods~\cite{sarafianos2024garment3dgen}.
Trained on large-scale 3D garment datasets~\cite{black2023bedlam,zou2023cloth4d,deitke2023objaverse}, \ourmethod demonstrates superior performance on held-out 3D garment data as well as in-the-wild clothing images. Extensive experiments show that our method outperforms state-of-the-art 2D-to-3D garment reconstruction approaches in terms of geometric accuracy, visual fidelity, and cross-view consistency.

\begin{figure*}[t]
    \centering
    \includegraphics[width=1\textwidth]{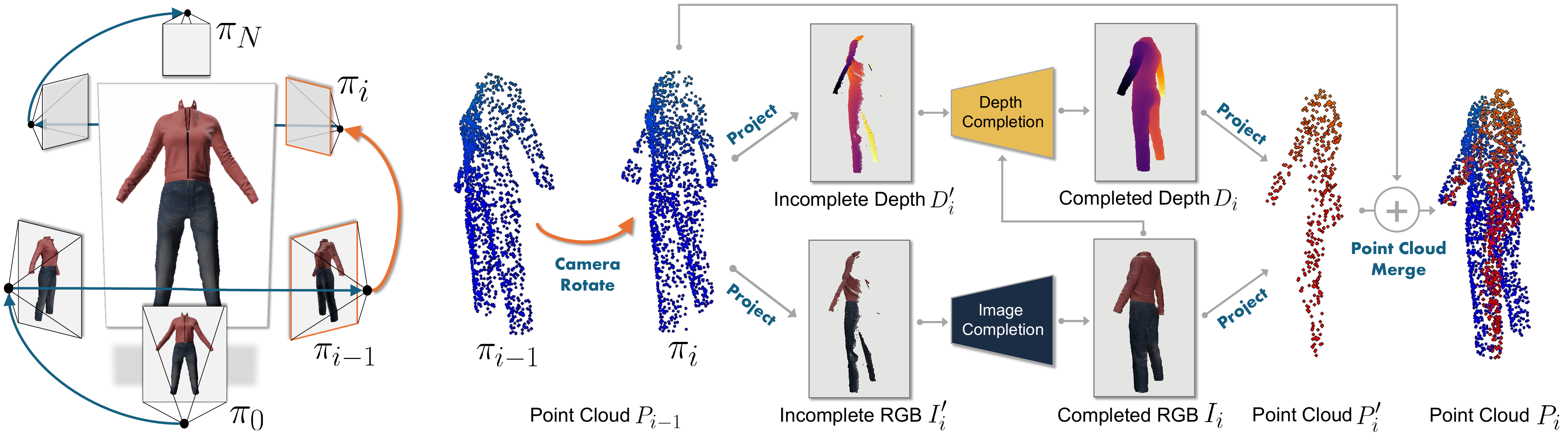}
    \vspace{-6.6mm}
    \caption{\small \textbf{An illustration of progressive novel view synthesis in \ourmethod.} \textbf{Left:} Given a garment image, our method performs depth-aware novel view synthesis along a predefined zigzag camera trajectory. \textbf{Right:} For each camera rotation from $\pi_{i-1}$ to $\pi_{i}$, we project the current point cloud $P_{i-1}$ into the image space based on camera pose $\pi_{i}$, resulting in incomplete RGB and depth images. Our diffusion model completes the RGB image using the warped view, input image, and camera pose as conditions, while a depth completion network refines the depth map based on the completed RGB, warped depth, and camera pose. The re-projected point cloud $P'_{i}$ is then merged with $P_{i-1}$ to produce an updated point cloud $P_{i}$. This iterative process continues until a full 3D representation of the garment is achieved. 
    }
    \vspace{-3mm}
    \label{fig:framework}
\end{figure*}

\section{Related Work}
\label{sec:related}

\mypara{Single-View 3D Garment Reconstruction and Editing.}
Reconstructing 3D garments from a single image has been widely explored, with existing methods approaching the task from several perspectives. One line of work relies on parametric body templates, such as SMPL~\cite{bhatnagar2019multi,corona2021smplicit,moon20223d,jiang2020bcnet}, or employs 2D shape priors and keypoint-based techniques~\cite{zhou2013garment} to optimize garment structure. Another category of work uses explicit or implicit 3D parametric garment models~\cite{li2024garment,danvevrek2017deepgarment,bhatnagar2019multi,gao2023cloth2tex,majithia2022robust,luo2024garverselod,sarafianos2024garment3dgen,zhu2022registering} to capture garment shape and support pose-guided deformations. Additionally, some methods incorporate garment sewing patterns~\cite{berthouzoz2013parsing,jeong2015garment,yang2016detailed,chen2022structure,chen2024panelformer,liu2023towards,zou2023cloth4d}, offering flexibility by reconstructing garments from 2D panels. 
However, these works often struggle to capture diverse garment styles and fine surface details (e.g., wrinkles), and lack support for intuitive garment manipulation, such as modifying surface details or garment parts. 
In contrast, \ourmethod prioritizes novel view synthesis for detailed geometry and texture reconstruction, without relying on garment templates or human body priors, allowing it to handle a wide range of garment styles. Furthermore, single-view edits can also be seamlessly extended to the 3D model. Note that, our focus in this paper is on garments in a rest pose --- well suited to the fashion industry, where ease of adjustment is essential.

\mypara{Novel View Synthesis from Sparse Images.}
Our method is inspired by novel view synthesis. Popular approaches such as Neural Radiance Fields (NeRFs)~\cite{mildenhall2020nerf} and 3D Gaussian Splatting (3D-GS)~\cite{kerbl20233d} rely on numerous posed inputs, limiting their use in single-view scenarios. Recently, distillation from pre-trained 2D generative models has emerged as a promising solution for hallucinating novel views from limited input, with applications in human digitization~\cite{saito2019pifu,zheng2021pamir,xiu2022icon,xiu2023econ,ho2024sith,kocabas2024hugs,alldieck2022photorealistic,hema2024famous} and object-centric reconstruction~\cite{watson2022novel,shi2023mvdream,hu2024mvd,poole2022dreamfusion,hu2024mvd,tang2023dreamgaussian,liu2023zero,liu2023syncdreamer,liu2024one,zhou2023sparsefusion}. However, these methods often lack cross-view consistency and high-quality details, crucial for garment-focused tasks.
Unlike models that sample views independently, our method takes semantic cues (i.e., wrapped images) from other views as an additional condition for view synthesis. This might be reminiscent of scene-level approaches, such as Perpetual View Synthesis~\cite{kaneva2010infinite,liu2021infinite,tung2025megascenes,yu2024wonderjourney,chung2023luciddreamer,cai2023diffdreamer}, which condition on warped images for neighbor view image completion. However, we note that scene-centric methods often lack the precision needed for object-centric cases (e.g., garment manipulation) and overlook loop closure for garment shape completion. Our work represents a novel attempt of progressive view synthesis with a predefined camera trajectory for garment reconstruction and editing.

\mypara{Image-to-3D Reconstruction.}
Our approach builds on recent advancements in image-to-3D reconstruction, where most methods distill pre-trained generative models via per-scene optimization~\cite{poole2022dreamfusion,wang2023score,chen2023fantasia3d,song2019generative,kwon2024generalizable} or multi-view diffusion techniques~\cite{liu2023zero,shi2023zero123++,liu2024one,liu2023syncdreamer,zhou2023sparsefusion,hu2024mvd,voleti2025sv3d}. With the availability of large-scale 3D datasets~\cite{deitke2023objaverse,deitke2024objaverse}, Large Reconstruction Models (LRMs)~\cite{xu2024instantmesh,hong2023lrm,li2023instant3d,tang2025lgm,xu2024grm} are being trained for feed-forward image-to-3D generation.
Unlike Zero-1-to-3 and its variants~\cite{liu2023zero}, our method leverages diffusion models to progressively condition on warped images with carefully designed camera trajectory and error reduction methods to enhance cross-view consistency. Additionally, we curated a 3D garment dataset, incorporating assets from existing 3D collections~\cite{deitke2023objaverse,black2023bedlam,zou2023cloth4d}, allowing our model to synthesize highly detailed, multi-view images and corresponding depth maps. This process yields multi-view images alongside accurate depth maps, enabling high-quality mesh reconstruction through standard point cloud-to-mesh methods~\cite{kazhdan2006poisson}. While we demonstrate point aggregation and mesh reconstruction in our work, our primary focus is on advancing the multi-view and depth synthesis stages rather than optimizing the point-to-mesh conversion process itself.

\section{Approach}
\label{sec:approach}
We first present problem statement in Section~\ref{sec:problem}, followed by our proposed progressive novel view synthesis in Section~\ref{sec:pnvs}. We introduce garment-centric applications enabled by our method in Section~\ref{sec:application}. We describe the details of data curation and model training methods in Section~\ref{sec:data}.

\subsection{Problem Definition}\label{sec:problem}

Given a single-view garment image $I_0$, our goal is to generate consistent novel views with detailed RGB textures and accurate depths, which support both single-view 3D reconstruction and editing. 
Specifically, we first estimate a depth map $D_0$ based on the input $I_0$. Then, we project every pixel in the foreground of the garment to the world space, creating a colored point cloud $P_0$. Our goal is to complete this point cloud by sequentially incorporating information from synthesized novel views. To achieve this, we propose an progressive 3D completion process with a predefined camera trajectory $\boldsymbol{\pi} = \{\pi_1, \pi_2, ... , \pi_N\}$ that forms a closed loop around the garment object. Figure~\ref{fig:framework} illustrates the overall framework. Next, we elaborate the details of an arbitrary step in the following sections.

\subsection{Progressive Novel View Synthesis}\label{sec:pnvs}

\mypara{Overview.} 
At the step $i$ of the progressive novel view synthesis (see Figure~\ref{fig:framework}), we first project the existing point cloud $P_{i-1}$ to the image plane of camera $\pi_{i} \in \boldsymbol{\pi}$, producing an incomplete image $I_i'$ and an incomplete depth map $D_i'$. We then apply an image completion model to inpaint the missing areas in $I_i'$, resulting in $I_i$. Next, we use an monocular depth estimation model to estimate the corresponding depth map $D_i$ consistent with the known depths in $D_i'$. Finally, we integrate $I_i$ and $D_i$ with the existing point cloud to obtain a merged $P_{i}$. By following a predefined camera trajectory, our method can generate view-dependent images and corresponding depths that enable high-quality garment reconstruction and edit with improved cross-view consistency.

\mypara{Conditional Image Generation.}
At step $i$, the goal is to synthesize $I_i \in \mathbb{R}^{H \times W \times 3}$, the image of the garment object from the viewpoint of camera $\pi_{i}$, given the input image $I_0$, the projected image $I_i'$, and the relative camera rotation $R_i \in \mathbb{R}^{3 \times 3}$ and translation $T_i \in \mathbb{R}^{3}$ from $\pi_{0}$ to $\pi_{i}$. We aim to train a model $f_\text{img}$ such that:
\begin{align}\label{eq:img}
    I_i = f_\text{img}(I_0, I_i', R_i, T_i), 
\end{align}
where $I_i$ is the synthesized complete image that retains the appearance of $I_i'$ in the known regions, and synthesizes plausible appearance in the unknown regions that remain perceptually consistent with $I_i'$ and the original input $I_0$. 

To learn $f_\text{img}$, we fine-tune a denoising diffusion model, leveraging its strong generalization capabilities in image generation. Specifically, we adopt a latent diffusion architecture based on Stable Diffusion~\cite{rombach2022high} with an image encoder $\mathcal{E}$, a denoising network $\epsilon_{\theta}$, and a decoder $\mathcal{D}$. At denoising step $s \in S$, let $z_s$ denote the noisy latent of the target image $x=I_i$, and let $\boldsymbol{c} = c(I_0, I_i', R_i, T_i)$ be the embedding of the anchor view image, target view projected image, and relative camera extrinsics. We optimize the following latent diffusion objective:
\begin{equation}
\mathcal{L}(\theta) = \mathbb{E}_{\mathcal{E} (I_0), \mathcal{E} (I_i'), \epsilon \sim \mathcal{N}(0, \mathbf{I}), s} \left[ \left\| \epsilon - \epsilon_{\theta}({z}_s, s, \boldsymbol{c} ) \right\|^2 \right].
\label{eq:diffusion_loss}
\end{equation}

Unlike existing multi-view diffusion models (e.g., \cite{liu2023zero,shi2023zero123++}), which synthesize novels views from an arbitrary input viewpoint, we unify our garment-centric task by fixing the input image to a near-frontal view of the garment. This allows ${R_i}$ and ${T_i}$ to be interpreted as the absolute camera transformation from the frontal view. Furthermore, in addition to conditioning on the anchor view image, we incorporate the warped image (i.e., $I_i'$ in Figure~\ref{fig:framework} and Equation~\ref{eq:img}) at the target view as an additional condition input, which provides a strong prior that enhances cross-view consistency in garment reconstruction, as demonstrated in Section~\ref{sec:ablation}.

\mypara{Conditional Depth Generation.} 
After obtained complete RGB image $I_i$, we learn a depth model $f_\text{depth}$ to estimate the depth map $D_i \in \mathbb{R}^{H \times W \times 1}$ conditioned on the warped incomplete depth map $D_i'$ as follows:
\begin{align}
D_i = f_\text{depth} (I_i, D_i')
\end{align}
Similar to the conditional image generation, we enforce depth preservation in known regions by framing the task as metric depth estimation. To ensure consistency, we align the depth values of $D_i$ and $D_i'$ during training. The model is optimized using an $\mathcal{L}_1$ loss:
\begin{align}
\mathcal{L}_1 = ||  (D_i - \hat{D_i}) \cdot m ||,
\end{align}
where $\hat{D_i}$ is the ground-truth depth, and $m$ is the foreground mask. To train $f_\text{depth}$, we fine-tune the pretrained human foundation model, Sapiens~\cite{khirodkar2025sapiens}, leveraging its strong priors for human-related tasks. To condition the model on $D_i'$, we concatenate $D_i'$ with $I_i$ as input and add an extra channel to the first projection layer of Sapiens model. The weights of the added channel are initialized to zero.

\mypara{Point Cloud Merging and Projection.}
To integrate novel view observations  (i.e., $I_i$ and $D_i$) into the existing point cloud $P_{i-1}$, we first identify the inpainted regions from the image model. Pixels in these regions are projected into world space and merged with $P_{i-1}$ to form $P_i$, with expanded borders to include overlapping regions. To minimize stitching artifacts, we align the depth map of the inpainted regions with the warped depth map of $P_{i-1}$.
When projecting a partial point cloud to a novel view, only surfaces facing the camera should be rendered. To enforce this, we track the orientation of each point. For a point $x$ added at step $i$, its orientation vector $v$ is derived from the normal direction of the corresponding pixel in $D_i$. During projection, a point is ignored if $\text{dot}~(v, v_0) < 0$, where $v_0$ is the viewing direction. After completing all steps along the camera trajectory, we optionally sample a few random views for additional inpainting to recover any occluded regions.
Please see supplementary for additional details.

\subsection{Garment Digitization and Editing}\label{sec:application}

\mypara{Garment Digitization.}
Our method enables garment digitization from a single image by progressively synthesizing novel views, generating multi-view consistent RGBD images and a colored point cloud. This output serves as an intermediate representation for various 3D reconstruction. In this work, we employ Screened Poisson surface reconstruction~\cite{kazhdan2006poisson} to convert the point cloud into a textured mesh. Specifically, we project multi-view RGBD images to form a colored point cloud, where each point encodes geometry and color. The Screened Poisson method then interpolates these attributes, mapping textures onto mesh vertices.

\mypara{Interactive Editing.}
Redesigning a 3D garment model typically requires significant expertise, making it impractical for most users. \ourmethod provides an intuitive alternative, allowing users to edit a rendered image of the garment from a selected view, which is then lifted into 3D. In this work, we focus on two types of edits: (1) \emph{Part-based Editing}: Modifies the geometry or texture of specific garment parts, such as sleeves or pant legs. Users can add, remove, or resize components. (2) \emph{Local surface editing}: Adjusts the geometry and texture of localized regions, such as adding a pocket or modifying the neckline design.

The garment part editing is achieved with the following strategy. Given a 3D garment object $G$, the user selects an anchor view $\pi$ and edits the rendered image $I$ to obtain $I_{\text{edit}}$. We first identify the edited region in $I_{\text{edit}}$ and remove the corresponding garment parts from $G$, leaving a partial garment $G'$ that remains unchanged. This reformulates the task as single-view 3D garment part reconstruction, conditioned on $G'$. We then follow the process described in Section~\ref{sec:pnvs} with two modifications: (1) At each step along the camera trajectory, the conditional image and depth are generated by combining the projected point cloud with observations from the partial garment $G'$. (2) After computing image and depth maps, only pixels within the edited region are projected and merged with the existing point cloud. The final output is a colored point cloud of the edited parts, which is then merged with $G'$. For local surface editing, instead of removing and reconstructing an entire garment part, we apply the same process to a localized surface region.

\subsection{Data Preparation and Training}\label{sec:data}
We construct the training dataset by simulating inference. For each 3D garment, we sample $6$ uniform views at $0^{\circ}$ elevation (following the full camera trajectory) and $4$ additional random views between $60^{\circ}$ and $-30^{\circ}$ for inpainting.

\mypara{Training Data for Reconstruction.}
We follow the zigzag camera trajectory (Figure~\ref{fig:framework}) and at each step $i$, we form a training pair for the image generation model $f_\text{img}$: $\{ (I_i', I_0, R_i, T_i), I_i \}$, where $I_i'$ is the projected image, $I_0$ is the anchor view, and $(R_i, T_i)$ are the relative camera transformations. Similarly, the depth generation model $f_\text{depth}$ is trained with $\{ (D_i', I_i), D_i \}$, where $D_i'$ is the projected depth, and $D_i$ is the ground-truth depth. We merge the point cloud with $I_i$ and $D_i$ before proceeding. Finally, we repeat the process for four random views to simulate inpainting.

\mypara{Training Data for Editing.}
For 3D editing, we generate training data by randomly removing parts of a 3D garment to create a partial known model. At each step, we create a partial image $I_i''$ and depth map $D_i''$ by merging $I_i'$ and $D_i'$ with known observations. The training pairs become $
\{(I_i'', I_0, R_i, T_i), I_i\}$ for $f_\text{img}$ and $\{(D_i'', I_i), D_i\}$ for $f_\text{depth}$.

\mypara{Joint Training.}
To learn a unified model for both reconstruction and editing, we combine their training data. We randomly apply small rotations to the 3D object when generating the training data, enabling the model to handle in-the-wild inputs that may not be well-posed. Please refer to the supplementary materials for details.

\section{Experiments}
\label{sec:exp}

We present experimental results of our method on single-view garment reconstruction and editing. Please see supplementary for additional details, analyses, and results.

\subsection{Datasets, Metrics, and Baselines}
\mypara{Datasets.}
We validate \ourmethod using 3D garment assets from a number of sources.
(1) Curated dataset: We collect $\sim$700 3D garments with diverse shape and texture from Artstation\footnote{https://www.artstation.com/}.
(2) Objaverse 1.0 (Garment)~\cite{deitke2023objaverse}: the original v1.0 dataset contains more than 800K 3D objects, where most of the existing method trained on~\cite{liu2023zero,xu2024instantmesh,yang2024hunyuan3d}. We manually curated a subset only contain $\sim$900  high-quality garment assets. 
(3) BEDLAM~\cite{black2023bedlam}: 114 garments, each has many textures, $\sim$1600 garments in total.
(4) Cloth4D~\cite{zou2023cloth4d}: $\sim$1100 artists made garments.

\mypara{Quantitative Metrics.}
(1) Texture and appearance quality: we evaluate the novel view synthesis using commonly used LPIPS~\cite{zhang2018unreasonable}, PSNR~\cite{hore2010image}, SSIM~\cite{wang2004image}.
(2) Geometry quality: we measure the performance using geometric errors with Chamfer distance (bi-directional point-to-mesh) between ground-truth and reconstructed meshes.

\mypara{Baselines.}
We compare \ourmethod with state-of-the-art models for image-to-3D object and image-to-garment reconstruction.
(1) InstantMesh~\cite{xu2024instantmesh}: object reconstruction by generating novel views using Zero-1-to-3++~\cite{shi2023zero123++}.
(2) CRM~\cite{wang2025crm}: generate six orthographic views for 3D object reconstruction.
(3) Hunyuan3D-1.0~\cite{yang2024hunyuan3d}: a newly released model for high-quality image-to-3D object reconstruction.
(4) Garment3DGen~\cite{sarafianos2024garment3dgen}: a state-of-the-art garment-specific model based on template optimization, with templates initialized by InstantMesh~\cite{xu2024instantmesh}. As the texture code is not released, we compare only mesh geometry.

\begin{figure*}[t]
    \centering
    \includegraphics[width=1\textwidth]{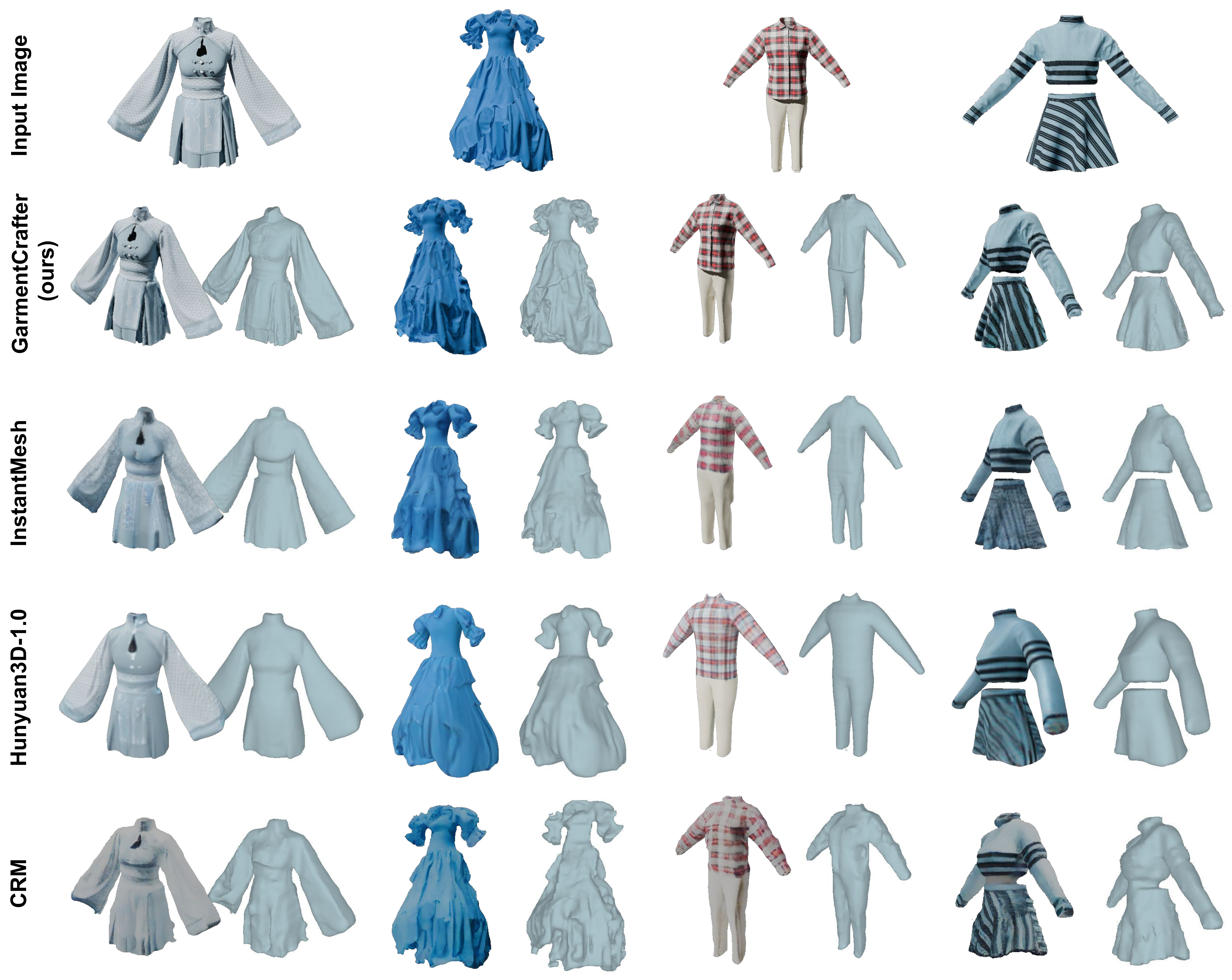}
    \caption{\small \textbf{Qualitative comparison on single-view 3D garment reconstruction with state-of-the-art methods.} Our method demonstrates  better performance in handling complex texture patterns and geometric structures compared to InstantMesh~\cite{xu2024instantmesh}, Hunyuan3D-1.0~\cite{yang2024tencent}, and Convolutional Reconstruction Model (CRM)~\cite{wang2025crm}.}
    \label{fig:sota_recon}
\end{figure*}

\begin{figure*}[t]
    \centering
    \includegraphics[width=1\textwidth]{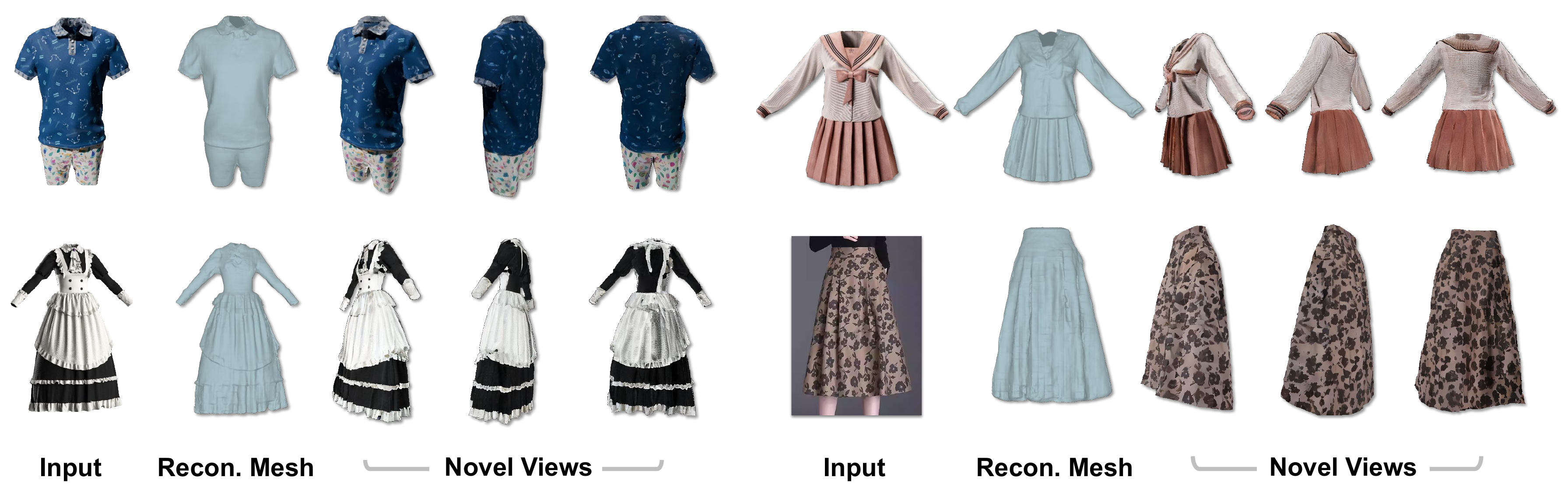}
    \caption{\small More qualitative results of \ourmethod on single-view reconstruction. Please see supplementary for more results.}
    \label{fig:recon}
\end{figure*}

\subsection{Results on Single-View Reconstruction}

We evaluate \ourmethod on single-view reconstruction using a held-out test dataset of 150 garment assets. For each test case, we sample 12 views with alternating elevations of 0$^\circ$ and 20$^\circ$ and azimuth angles evenly spaced over 360$^\circ$. To assess image quality, we convert the generated point clouds to meshes using a classical surface reconstruction method and render multi-view images. For geometry evaluation, we compute the Chamfer distance directly between the generated point cloud and the ground-truth mesh.

\mypara{Qualitative Results.}
Figure~\ref{fig:sota_recon} shows qualitative comparisons, where \ourmethod demonstrates superior texture and geometry generation compared to all other baselines. Our method, benefiting from consistent multi-view generation, produces sharp textures and intricate geometric details, whereas other baselines often result in blurry textures and overly smoothed geometries. Figure~\ref{fig:recon} shows additional qualitative results of \ourmethod.

\mypara{Quantitative Results on Texture Quality.}
We conduct a quantitative analysis of texture quality on our held-out test dataset and show results in Table~\ref{tab:sota_2d_3d}. Across all image quality metrics, \ourmethod consistently surpasses baseline methods, demonstrating its effectiveness in producing high-fidelity textures and preserving fine-grained details.

\begin{table}[t]
\small
\tabcolsep 4pt 
\centering
\caption{\small \textbf{Quantitative comparison of texture and geometry quality.} InstantMesh${\star}$: with fine-tuned Zero-1-to-3++ on our garment data for a fair comparison. CRM and Hunyuan3D-1.0 require significant computing for full fine-tuning, making it impractical. Garment3DGen does not provide texture reconstruction code.}
\begin{tabular}{l|ccc|c}
\toprule
 & \multicolumn{3}{c|}{Appearance} & \multicolumn{1}{c}{Geometry} \\
 \cmidrule{2-5}
 & \multicolumn{1}{c}{LPIPS$\downarrow$} & \multicolumn{1}{c}{PSNR$\uparrow$} & \multicolumn{1}{c|}{SSIM$\uparrow$} & \multicolumn{1}{c}{Chamfer$\downarrow$} \\
  \midrule
InstantMesh${\star}$~\cite{xu2024instantmesh}  & \underline{0.1848} & \underline{19.14} & 0.7944 &  0.0139 \\
CRM~\cite{wang2025crm}  & 0.2213 & 17.51 & \underline{0.8131} & 0.0127 \\
Hunyuan3D-1.0~\cite{yang2024hunyuan3d}  & 0.2216 & 17.77 & 0.7794 & \underline{0.0121} \\
Garment3DGen~\cite{sarafianos2024garment3dgen}  & -- & -- & -- &  0.0123 \\
\midrule
\textbf{\ourmethod}  & \textbf{0.1190} & \textbf{22.36} & \textbf{0.8317} & \textbf{0.0044} \\
\bottomrule
\end{tabular}
\label{tab:sota_2d_3d}
\end{table}

\mypara{Quantitative Results on Geometry Quality.}
We present quantitative geometry evaluation results in Table~\ref{tab:sota_2d_3d}. \ourmethod outperforms baseline methods in terms of Chamfer distance, highlighting its enhanced ability to capture detailed surface geometries in 3D garment shapes.

\subsection{Results on Single-View Editing}

We present qualitative results on single-view editing in Figure~\ref{fig:edit}, showcasing various types of edits, including resizing, element swapping, and surface editing. \ourmethod successfully applies 3D edits that are consistent with the 2D edits, while preserving cross-view consistency.

\subsection{Analyses and Ablation Studies}\label{sec:ablation}

\mypara{Importance of Progressive Novel View Synthesis.}
A key insight of our method is to progressively synthesize novel view by conditioning the generation on the projected images. We conduct an ablation study on the effect of projected image  conditioning.
For each test case, we select an anchor view $\pi_1$, and a second camera view, $\pi_2$, at a 60° azimuthal angle relative to $\pi_1$. 
We compare the performance of our image model with or without projected image conditioning at synthesizing view $\pi_2$ in Table~\ref{tab:ablation_pnvs}. We observe a drop in performance measured in image similarity metrics when removing the projected condition.

\begin{table}[t] \label{consistency table}
\small
\tabcolsep 9pt 
\centering
\caption{\small Ablation study on Progressive Novel View Synthesis (P-NVS) and analysis on multi-view consistency. We show results with and without P-NVS. CVCS: Cross-View Consistency Score.}
\begin{tabular}{c|ccc|c}
\toprule
P-NVS & LPIPS $\downarrow$ & PSNR $\uparrow$ & SSIM $\uparrow$ & CVCS$\uparrow$ \\
\midrule
\xmark & 0.1195 & 21.512 & 0.8369 & 0.9030 \\
\cmark & \textbf{0.1052} & \textbf{22.776} & \textbf{0.8557} & \textbf{0.9512} \\
\bottomrule
\end{tabular}\label{tab:ablation_pnvs}
\end{table}

\begin{figure}[t]
    \centerline{\includegraphics[width=0.95\linewidth]{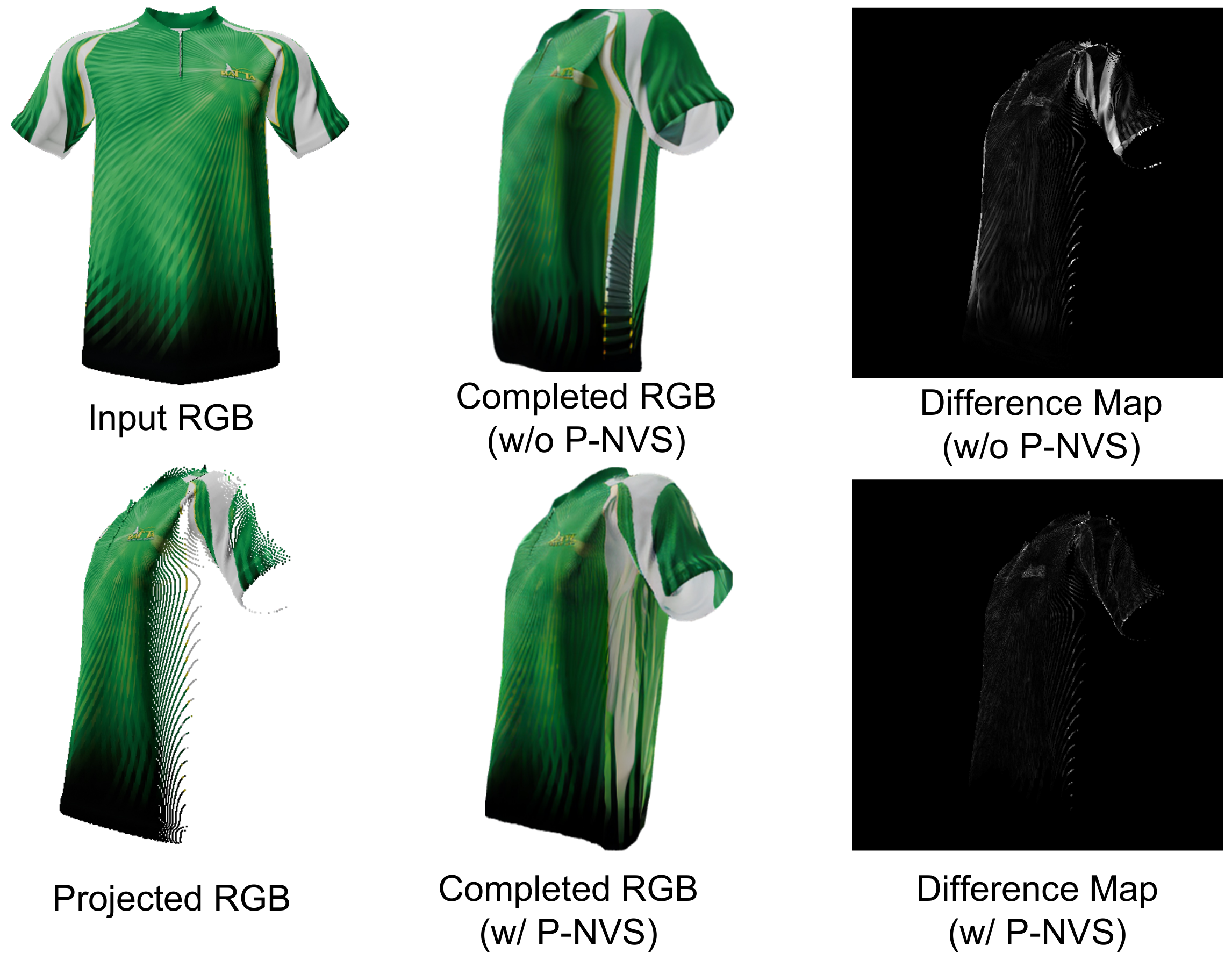}}
    \caption{\small \textbf{Analysis of projected image conditioning.} Left: we show original input and projected RGB images. Middle: completed RGB images with and without Progressive Novel View Synthesis (P-NVS). Right: difference between completed and projected images, showing our novel view aligns more closely with the ground-truth projected RGB. Zoom-in for details.}
  \label{fig:diff}
\end{figure}

\begin{figure*}[t]
    \centering
    \includegraphics[width=1\textwidth]{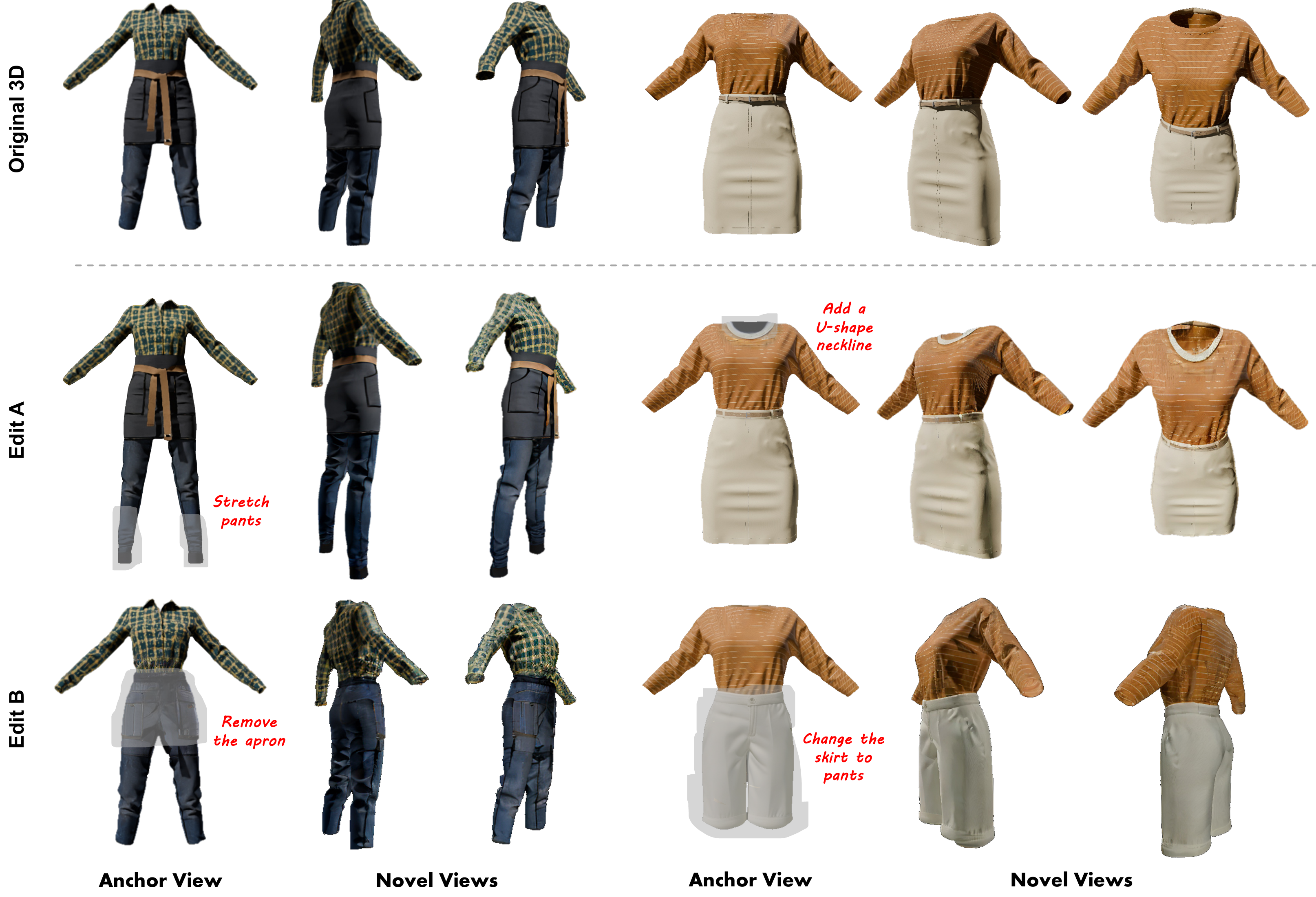}
    \vspace{-6mm}
    \caption{\small \textbf{Qualitative results on single-view 3D garment editing.} \ourmethod enables single-view edit such as modify the geometry and surface details of the garment, with the changes accurately reflected across the 3D model. Please see supplementary for more results.
    }\vspace{-3mm}
    \label{fig:edit}
\end{figure*}

\mypara{Analysis on multi-view consistency.}
Common image metrics (e.g., LPIPS, PSNR, and SSIM) measure similarity but do not directly reflect cross-view consistency. To address this, we propose a new metric, the Cross-View Consistency Score (CVCS), to gain deeper insights into the consistency performance of our model.
\begin{equation}
    \text{CVCS} = 1 - \frac{\Sigma |I-I'| \cdot m'}{\Sigma m'}
\end{equation}
where $I$ is the synthesized image at camera view $\pi$, $I'$ is a partial image projected from an observed view $\pi_0$ with known depth, and $m'$ is a binary mask indicating the projection regions. This assumes $\pi$ and $\pi_0$ are relatively close.

We use the CVCS metric to ablate the impact of P-NVS.  As shown in Table \ref{tab:ablation_pnvs},  \ourmethod achieves superior cross-view consistency with P-NVS. We further validates this claim with a visual example in Figure \ref{fig:diff}. While both model synthesizes plausible novel views, \ourmethod with P-NVS aligns more closely with the input observation.

\begin{figure}[t]
    \centerline{\includegraphics[width=1\linewidth]{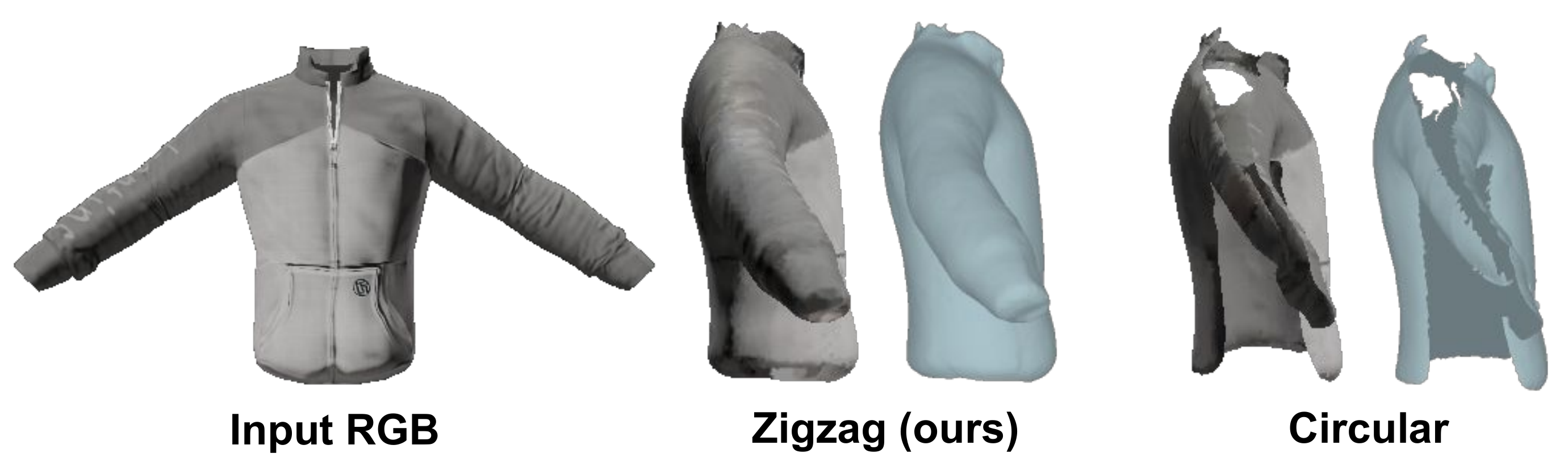}}
    \vspace{-2mm}
    \caption{\small \textbf{Camera trajectory selection for loop closure.} Zigzag achieves better loop closure, while the circular trajectory struggles with side-view closure, leading to geometric conflicts and model failure. We argue that there are numerous ways to select camera trajectories, our proposed approach just offers an intuitive solution tailored for single-view garment reconstruction and editing.}
    \label{fig:trajectory}
\end{figure}

\mypara{Effect of Trajectory on Loop Closure.}
For better loop closure, we use a ``zigzag'' camera trajectory where we rotate the camera to left and right alternatively and converge at the center back of the garment (see Figure~\ref{fig:framework}). This design aims to better capture overlapping views, thereby improving reconstruction accuracy. We validate this design choice by comparing the quality of the 3D meshes generated using zigzag and sequential trajectories. We report quantitative results in Table~\ref{tab:ablation_trajectory}. We find that our chosen trajectory achieves better performance across both image and geometry metrics.
We additionally show a qualitative comparison in Figure~\ref{fig:trajectory}. When using a circular trajectory, achieving loop closure from the side view is challenging; the generated geometry (left sleeve) often conflicts with prior predictions, leading to model failure.

\begin{table}[t]
\small
\tabcolsep 6pt 
\centering
\caption{\small \textbf{Ablation study on camera trajectory selection.} We study two types camera trajectory for progressive novel view synthesis. \textbf{Circular}: the camera moves around the object in regular steps, either clockwise or counterclockwise. \textbf{Zigzag}: the camera alternates directions with each step, as shown in Figure~\ref{fig:framework}. Results indicate that our proposed zigzag achieves better appearance and geometry quality compared to using circular trajectory. We show an actual example in Figure~\ref{fig:trajectory} for qualitative analyses.}
\vspace{-2mm}
\begin{tabular}{l|ccc|c}
\toprule
Trajectory &  LPIPS $\downarrow$ & PSNR $\uparrow$ & SSIM $\uparrow$ & Chamfer $\downarrow$\\
 \midrule
Circular & 0.1503 & 20.79 & 0.8130 & 0.0054 \\
Zigzag (ours)  & \textbf{0.1454} & \textbf{21.22} & \textbf{0.8173} & \textbf{0.0044} \\
\bottomrule
\end{tabular}\label{tab:ablation_trajectory}
\end{table}

\section{Conclusion}
\label{sec:conclusion}

We present \ourmethod, a new approach to reconstruct and edit 3D garments from a single input image. Our method synthesizes novel view images progressively to ensure cross-view consistency, thereby achieving high quality geometry and texture results. We have conducted extensive experiments to demonstrate the superior performance of \ourmethod with other baseline methods.
Please see supplementary materials for additional implementation and training details, more qualitative results on garment reconstruction and editing, as well as an ablation study on the rotation angles in the camera trajectory.

\mypara{Limitation and future works.}
We focus on garments in a rest pose and cannot handle arbitrary poses. In addition, our model reconstructs only the external surface, not inner layers or structures. These will be addressed in future work.

{
    \small
    \bibliographystyle{ieeenat_fullname}
    \bibliography{main}

\begin{thebibliography}{88}
\providecommand{\natexlab}[1]{#1}
\providecommand{\url}[1]{\texttt{#1}}
\expandafter\ifx\csname urlstyle\endcsname\relax
  \providecommand{\doi}[1]{doi: #1}\else
  \providecommand{\doi}{doi: \begingroup \urlstyle{rm}\Url}\fi

\bibitem[Clo()]{Clo3D}
{CLO3D}.
\newblock \url{https://www.clo3d.com/en/}.

\bibitem[Sty()]{Style3D}
{Style3D}.
\newblock \url{https://www.linctex.com/}.

\bibitem[TUK()]{TUKA3D}
{TUKA3D}.
\newblock \url{https://tukatech.com/tuka3d/}.

\bibitem[bro()]{browzwear}
{Browzwear}.
\newblock \url{https://browzwear.com/}.

\bibitem[Alldieck et~al.(2022)Alldieck, Zanfir, and Sminchisescu]{alldieck2022photorealistic}
Thiemo Alldieck, Mihai Zanfir, and Cristian Sminchisescu.
\newblock Photorealistic monocular 3d reconstruction of humans wearing clothing.
\newblock In \emph{Proceedings of the IEEE/CVF Conference on Computer Vision and Pattern Recognition}, pages 1506--1515, 2022.

\bibitem[Berthouzoz et~al.(2013)Berthouzoz, Garg, Kaufman, Grinspun, and Agrawala]{berthouzoz2013parsing}
Floraine Berthouzoz, Akash Garg, Danny~M Kaufman, Eitan Grinspun, and Maneesh Agrawala.
\newblock Parsing sewing patterns into 3d garments.
\newblock \emph{Acm Transactions on Graphics (TOG)}, 32\penalty0 (4):\penalty0 1--12, 2013.

\bibitem[Bhatnagar et~al.(2019)Bhatnagar, Tiwari, Theobalt, and Pons-Moll]{bhatnagar2019multi}
Bharat~Lal Bhatnagar, Garvita Tiwari, Christian Theobalt, and Gerard Pons-Moll.
\newblock Multi-garment net: Learning to dress 3d people from images.
\newblock In \emph{Proceedings of the IEEE/CVF international conference on computer vision}, pages 5420--5430, 2019.

\bibitem[Black et~al.(2023)Black, Patel, Tesch, and Yang]{black2023bedlam}
Michael~J Black, Priyanka Patel, Joachim Tesch, and Jinlong Yang.
\newblock Bedlam: A synthetic dataset of bodies exhibiting detailed lifelike animated motion.
\newblock In \emph{Proceedings of the IEEE/CVF Conference on Computer Vision and Pattern Recognition}, pages 8726--8737, 2023.

\bibitem[Brooks et~al.(2023)Brooks, Holynski, and Efros]{brooks2023instructpix2pix}
Tim Brooks, Aleksander Holynski, and Alexei~A Efros.
\newblock Instructpix2pix: Learning to follow image editing instructions.
\newblock In \emph{Proceedings of the IEEE/CVF Conference on Computer Vision and Pattern Recognition}, pages 18392--18402, 2023.

\bibitem[Cai et~al.(2023)Cai, Chan, Peng, Shahbazi, Obukhov, Van~Gool, and Wetzstein]{cai2023diffdreamer}
Shengqu Cai, Eric~Ryan Chan, Songyou Peng, Mohamad Shahbazi, Anton Obukhov, Luc Van~Gool, and Gordon Wetzstein.
\newblock Diffdreamer: Towards consistent unsupervised single-view scene extrapolation with conditional diffusion models.
\newblock In \emph{Proceedings of the IEEE/CVF International Conference on Computer Vision}, pages 2139--2150, 2023.

\bibitem[Casado-Elvira et~al.(2022)Casado-Elvira, Trinidad, and Casas]{casado2022pergamo}
Andr{\'e}s Casado-Elvira, Marc~Comino Trinidad, and Dan Casas.
\newblock Pergamo: Personalized 3d garments from monocular video.
\newblock In \emph{Computer Graphics Forum}, pages 293--304. Wiley Online Library, 2022.

\bibitem[Chen et~al.(2024)Chen, Su, Hu, Yao, and Chu]{chen2024panelformer}
Cheng-Hsiu Chen, Jheng-Wei Su, Min-Chun Hu, Chih-Yuan Yao, and Hung-Kuo Chu.
\newblock Panelformer: Sewing pattern reconstruction from 2d garment images.
\newblock In \emph{Proceedings of the IEEE/CVF Winter Conference on Applications of Computer Vision}, pages 454--463, 2024.

\bibitem[Chen et~al.(2023)Chen, Chen, Jiao, and Jia]{chen2023fantasia3d}
Rui Chen, Yongwei Chen, Ningxin Jiao, and Kui Jia.
\newblock Fantasia3d: Disentangling geometry and appearance for high-quality text-to-3d content creation.
\newblock In \emph{Proceedings of the IEEE/CVF international conference on computer vision}, pages 22246--22256, 2023.

\bibitem[Chen et~al.(2022)Chen, Wang, Zhu, Liang, Torr, and Lin]{chen2022structure}
Xipeng Chen, Guangrun Wang, Dizhong Zhu, Xiaodan Liang, Philip Torr, and Liang Lin.
\newblock Structure-preserving 3d garment modeling with neural sewing machines.
\newblock \emph{Advances in Neural Information Processing Systems}, 35:\penalty0 15147--15159, 2022.

\bibitem[Chung et~al.(2023)Chung, Lee, Nam, Lee, and Lee]{chung2023luciddreamer}
Jaeyoung Chung, Suyoung Lee, Hyeongjin Nam, Jaerin Lee, and Kyoung~Mu Lee.
\newblock Luciddreamer: Domain-free generation of 3d gaussian splatting scenes.
\newblock \emph{arXiv preprint arXiv:2311.13384}, 2023.

\bibitem[Corona et~al.(2021)Corona, Pumarola, Alenya, Pons-Moll, and Moreno-Noguer]{corona2021smplicit}
Enric Corona, Albert Pumarola, Guillem Alenya, Gerard Pons-Moll, and Francesc Moreno-Noguer.
\newblock Smplicit: Topology-aware generative model for clothed people.
\newblock In \emph{Proceedings of the IEEE/CVF conference on computer vision and pattern recognition}, pages 11875--11885, 2021.

\bibitem[Dan{\v{e}}{\v{r}}ek et~al.(2017)Dan{\v{e}}{\v{r}}ek, Dibra, {\"O}ztireli, Ziegler, and Gross]{danvevrek2017deepgarment}
R Dan{\v{e}}{\v{r}}ek, Endri Dibra, Cengiz {\"O}ztireli, Remo Ziegler, and Markus Gross.
\newblock Deepgarment: 3d garment shape estimation from a single image.
\newblock In \emph{Computer Graphics Forum}, pages 269--280. Wiley Online Library, 2017.

\bibitem[Deitke et~al.(2023)Deitke, Schwenk, Salvador, Weihs, Michel, VanderBilt, Schmidt, Ehsani, Kembhavi, and Farhadi]{deitke2023objaverse}
Matt Deitke, Dustin Schwenk, Jordi Salvador, Luca Weihs, Oscar Michel, Eli VanderBilt, Ludwig Schmidt, Kiana Ehsani, Aniruddha Kembhavi, and Ali Farhadi.
\newblock Objaverse: A universe of annotated 3d objects.
\newblock In \emph{Proceedings of the IEEE/CVF Conference on Computer Vision and Pattern Recognition}, pages 13142--13153, 2023.

\bibitem[Deitke et~al.(2024)Deitke, Liu, Wallingford, Ngo, Michel, Kusupati, Fan, Laforte, Voleti, Gadre, et~al.]{deitke2024objaverse}
Matt Deitke, Ruoshi Liu, Matthew Wallingford, Huong Ngo, Oscar Michel, Aditya Kusupati, Alan Fan, Christian Laforte, Vikram Voleti, Samir~Yitzhak Gadre, et~al.
\newblock Objaverse-xl: A universe of 10m+ 3d objects.
\newblock \emph{Advances in Neural Information Processing Systems}, 36, 2024.

\bibitem[Gao et~al.(2024)Gao, Chen, Zhang, Wang, Sun, Zhang, Bo, and Huang]{gao2023cloth2tex}
Daiheng Gao, Xu Chen, Xindi Zhang, Qi Wang, Ke Sun, Bang Zhang, Liefeng Bo, and Qixing Huang.
\newblock Cloth2tex: A customized cloth texture generation pipeline for 3d virtual try-on.
\newblock In \emph{Proceedings of the IEEE/CVF Winter Conference on Applications of Computer Vision (WACV)}, 2024.

\bibitem[Guo et~al.(2023)Guo, Jiang, Chen, Song, and Hilliges]{guo2023vid2avatar}
Chen Guo, Tianjian Jiang, Xu Chen, Jie Song, and Otmar Hilliges.
\newblock Vid2avatar: 3d avatar reconstruction from videos in the wild via self-supervised scene decomposition.
\newblock In \emph{Proceedings of the IEEE/CVF Conference on Computer Vision and Pattern Recognition}, pages 12858--12868, 2023.

\bibitem[Hema et~al.(2024)Hema, Aich, Haene, Bazin, and De~la Torre]{hema2024famous}
Vishnu~Mani Hema, Shubhra Aich, Christian Haene, Jean-Charles Bazin, and Fernando De~la Torre.
\newblock Famous: High-fidelity monocular 3d human digitization using view synthesis.
\newblock \emph{arXiv preprint arXiv:2410.09690}, 2024.

\bibitem[Ho et~al.(2024)Ho, Song, Hilliges, et~al.]{ho2024sith}
I Ho, Jie Song, Otmar Hilliges, et~al.
\newblock Sith: Single-view textured human reconstruction with image-conditioned diffusion.
\newblock In \emph{Proceedings of the IEEE/CVF Conference on Computer Vision and Pattern Recognition}, pages 538--549, 2024.

\bibitem[Hong et~al.(2023)Hong, Zhang, Gu, Bi, Zhou, Liu, Liu, Sunkavalli, Bui, and Tan]{hong2023lrm}
Yicong Hong, Kai Zhang, Jiuxiang Gu, Sai Bi, Yang Zhou, Difan Liu, Feng Liu, Kalyan Sunkavalli, Trung Bui, and Hao Tan.
\newblock {LRM}: Large reconstruction model for single image to 3d.
\newblock \emph{arXiv preprint arXiv:2311.04400}, 2023.

\bibitem[Hore and Ziou(2010)]{hore2010image}
Alain Hore and Djemel Ziou.
\newblock Image quality metrics: Psnr vs. ssim.
\newblock In \emph{2010 20th international conference on pattern recognition}, pages 2366--2369. IEEE, 2010.

\bibitem[Hu et~al.(2024{\natexlab{a}})Hu, Zhou, Jampani, and Tulsiani]{hu2024mvd}
Hanzhe Hu, Zhizhuo Zhou, Varun Jampani, and Shubham Tulsiani.
\newblock Mvd-fusion: Single-view 3d via depth-consistent multi-view generation.
\newblock In \emph{Proceedings of the IEEE/CVF Conference on Computer Vision and Pattern Recognition}, pages 9698--9707, 2024{\natexlab{a}}.

\bibitem[Hu et~al.(2024{\natexlab{b}})Hu, Zhang, Zhang, Zhou, Liu, Zhang, and Nie]{hu2024gaussianavatar}
Liangxiao Hu, Hongwen Zhang, Yuxiang Zhang, Boyao Zhou, Boning Liu, Shengping Zhang, and Liqiang Nie.
\newblock Gaussianavatar: Towards realistic human avatar modeling from a single video via animatable 3d gaussians.
\newblock In \emph{Proceedings of the IEEE/CVF Conference on Computer Vision and Pattern Recognition}, pages 634--644, 2024{\natexlab{b}}.

\bibitem[Jeong et~al.(2015)Jeong, Han, and Ko]{jeong2015garment}
Moon-Hwan Jeong, Dong-Hoon Han, and Hyeong-Seok Ko.
\newblock Garment capture from a photograph.
\newblock \emph{Computer Animation and Virtual Worlds}, 26\penalty0 (3-4):\penalty0 291--300, 2015.

\bibitem[Jiang et~al.(2020)Jiang, Zhang, Hong, Luo, Liu, and Bao]{jiang2020bcnet}
Boyi Jiang, Juyong Zhang, Yang Hong, Jinhao Luo, Ligang Liu, and Hujun Bao.
\newblock Bcnet: Learning body and cloth shape from a single image.
\newblock In \emph{Computer Vision--ECCV 2020: 16th European Conference, Glasgow, UK, August 23--28, 2020, Proceedings, Part XX 16}, pages 18--35. Springer, 2020.

\bibitem[Kaneva et~al.(2010)Kaneva, Sivic, Torralba, Avidan, and Freeman]{kaneva2010infinite}
Biliana Kaneva, Josef Sivic, Antonio Torralba, Shai Avidan, and William~T Freeman.
\newblock Infinite images: Creating and exploring a large photorealistic virtual space.
\newblock \emph{Proceedings of the IEEE}, 98\penalty0 (8):\penalty0 1391--1407, 2010.

\bibitem[Kazhdan et~al.(2006)Kazhdan, Bolitho, and Hoppe]{kazhdan2006poisson}
Michael Kazhdan, Matthew Bolitho, and Hugues Hoppe.
\newblock Poisson surface reconstruction.
\newblock In \emph{Proceedings of the fourth Eurographics symposium on Geometry processing}, 2006.

\bibitem[Kerbl et~al.(2023)Kerbl, Kopanas, Leimk{\"u}hler, and Drettakis]{kerbl20233d}
Bernhard Kerbl, Georgios Kopanas, Thomas Leimk{\"u}hler, and George Drettakis.
\newblock 3d gaussian splatting for real-time radiance field rendering.
\newblock \emph{ACM Transactions on Graphics}, 42\penalty0 (4):\penalty0 1--14, 2023.

\bibitem[Khirodkar et~al.(2024)Khirodkar, Bagautdinov, Martinez, Zhaoen, James, Selednik, Anderson, and Saito]{khirodkar2025sapiens}
Rawal Khirodkar, Timur Bagautdinov, Julieta Martinez, Su Zhaoen, Austin James, Peter Selednik, Stuart Anderson, and Shunsuke Saito.
\newblock Sapiens: Foundation for human vision models.
\newblock In \emph{European Conference on Computer Vision}, pages 206--228. Springer, 2024.

\bibitem[Kocabas et~al.(2024)Kocabas, Chang, Gabriel, Tuzel, and Ranjan]{kocabas2024hugs}
Muhammed Kocabas, Jen-Hao~Rick Chang, James Gabriel, Oncel Tuzel, and Anurag Ranjan.
\newblock Hugs: Human gaussian splats.
\newblock In \emph{Proceedings of the IEEE/CVF conference on computer vision and pattern recognition}, pages 505--515, 2024.

\bibitem[Kwon et~al.(2024)Kwon, Fang, Lu, Dong, Zhang, Carrasco, Mosella-Montoro, Xu, Takagi, Kim, et~al.]{kwon2024generalizable}
Youngjoong Kwon, Baole Fang, Yixing Lu, Haoye Dong, Cheng Zhang, Francisco~Vicente Carrasco, Albert Mosella-Montoro, Jianjin Xu, Shingo Takagi, Daeil Kim, et~al.
\newblock Generalizable human gaussians for sparse view synthesis.
\newblock In \emph{Proceedings of the European Conference on Computer Vision (ECCV)}, 2024.

\bibitem[Li et~al.(2023)Li, Tan, Zhang, Xu, Luan, Xu, Hong, Sunkavalli, Shakhnarovich, and Bi]{li2023instant3d}
Jiahao Li, Hao Tan, Kai Zhang, Zexiang Xu, Fujun Luan, Yinghao Xu, Yicong Hong, Kalyan Sunkavalli, Greg Shakhnarovich, and Sai Bi.
\newblock Instant3d: Fast text-to-3d with sparse-view generation and large reconstruction model.
\newblock \emph{arXiv preprint arXiv:2311.06214}, 2023.

\bibitem[Li et~al.(2024)Li, Dumery, Guillard, and Fua]{li2024garment}
Ren Li, Corentin Dumery, Beno{\^\i}t Guillard, and Pascal Fua.
\newblock Garment recovery with shape and deformation priors.
\newblock In \emph{Proceedings of the IEEE/CVF Conference on Computer Vision and Pattern Recognition}, pages 1586--1595, 2024.

\bibitem[Liu et~al.(2021)Liu, Tucker, Jampani, Makadia, Snavely, and Kanazawa]{liu2021infinite}
Andrew Liu, Richard Tucker, Varun Jampani, Ameesh Makadia, Noah Snavely, and Angjoo Kanazawa.
\newblock Infinite nature: Perpetual view generation of natural scenes from a single image.
\newblock In \emph{Proceedings of the IEEE/CVF International Conference on Computer Vision}, pages 14458--14467, 2021.

\bibitem[Liu et~al.(2023{\natexlab{a}})Liu, Xu, Lin, Liang, and Yan]{liu2023towards}
Lijuan Liu, Xiangyu Xu, Zhijie Lin, Jiabin Liang, and Shuicheng Yan.
\newblock Towards garment sewing pattern reconstruction from a single image.
\newblock \emph{ACM Transactions on Graphics (TOG)}, 42\penalty0 (6):\penalty0 1--15, 2023{\natexlab{a}}.

\bibitem[Liu et~al.(2024)Liu, Xu, Jin, Chen, Varma~T, Xu, and Su]{liu2024one}
Minghua Liu, Chao Xu, Haian Jin, Linghao Chen, Mukund Varma~T, Zexiang Xu, and Hao Su.
\newblock One-2-3-45: Any single image to 3d mesh in 45 seconds without per-shape optimization.
\newblock \emph{Advances in Neural Information Processing Systems}, 36, 2024.

\bibitem[Liu et~al.(2023{\natexlab{b}})Liu, Wu, Van~Hoorick, Tokmakov, Zakharov, and Vondrick]{liu2023zero}
Ruoshi Liu, Rundi Wu, Basile Van~Hoorick, Pavel Tokmakov, Sergey Zakharov, and Carl Vondrick.
\newblock Zero-1-to-3: Zero-shot one image to 3d object.
\newblock In \emph{Proceedings of the IEEE/CVF international conference on computer vision}, pages 9298--9309, 2023{\natexlab{b}}.

\bibitem[Liu et~al.(2023{\natexlab{c}})Liu, Lin, Zeng, Long, Liu, Komura, and Wang]{liu2023syncdreamer}
Yuan Liu, Cheng Lin, Zijiao Zeng, Xiaoxiao Long, Lingjie Liu, Taku Komura, and Wenping Wang.
\newblock Syncdreamer: Generating multiview-consistent images from a single-view image.
\newblock \emph{arXiv preprint arXiv:2309.03453}, 2023{\natexlab{c}}.

\bibitem[Loper et~al.(2023)Loper, Mahmood, Romero, Pons-Moll, and Black]{loper2023smpl}
Matthew Loper, Naureen Mahmood, Javier Romero, Gerard Pons-Moll, and Michael~J Black.
\newblock {SMPL}: A skinned multi-person linear model.
\newblock In \emph{Seminal Graphics Papers: Pushing the Boundaries, Volume 2}, pages 851--866. 2023.

\bibitem[Luo et~al.(2024)Luo, Liu, Li, Du, Jin, Sun, Nie, Chen, and Han]{luo2024garverselod}
Zhongjin Luo, Haolin Liu, Chenghong Li, Wanghao Du, Zirong Jin, Wanhu Sun, Yinyu Nie, Weikai Chen, and Xiaoguang Han.
\newblock Garverselod: High-fidelity 3d garment reconstruction from a single in-the-wild image using a dataset with levels of details.
\newblock \emph{arXiv preprint arXiv:2411.03047}, 2024.

\bibitem[Majithia et~al.(2022)Majithia, Parameswaran, Babar, Garg, Srivastava, and Sharma]{majithia2022robust}
Sahib Majithia, Sandeep~N Parameswaran, Sadbhavana Babar, Vikram Garg, Astitva Srivastava, and Avinash Sharma.
\newblock Robust 3d garment digitization from monocular 2d images for 3d virtual try-on systems.
\newblock In \emph{Proceedings of the IEEE/CVF Winter Conference on Applications of Computer Vision}, pages 3428--3438, 2022.

\bibitem[Mildenhall et~al.(2020)Mildenhall, Srinivasan, Tancik, Barron, Ramamoorthi, and Ng]{mildenhall2020nerf}
Ben Mildenhall, Pratul~P. Srinivasan, Matthew Tancik, Jonathan~T. Barron, Ravi Ramamoorthi, and Ren Ng.
\newblock Nerf: Representing scenes as neural radiance fields for view synthesis.
\newblock In \emph{European Conference on Computer Vision (ECCV)}, 2020.

\bibitem[Moon et~al.(2022)Moon, Nam, Shiratori, and Lee]{moon20223d}
Gyeongsik Moon, Hyeongjin Nam, Takaaki Shiratori, and Kyoung~Mu Lee.
\newblock 3d clothed human reconstruction in the wild.
\newblock In \emph{European conference on computer vision}, pages 184--200. Springer, 2022.

\bibitem[Pan et~al.(2023)Pan, Tewari, Leimk{\"u}hler, Liu, Meka, and Theobalt]{pan2023drag}
Xingang Pan, Ayush Tewari, Thomas Leimk{\"u}hler, Lingjie Liu, Abhimitra Meka, and Christian Theobalt.
\newblock Drag your gan: Interactive point-based manipulation on the generative image manifold.
\newblock In \emph{ACM SIGGRAPH 2023 Conference Proceedings}, pages 1--11, 2023.

\bibitem[Poole et~al.(2022)Poole, Jain, Barron, and Mildenhall]{poole2022dreamfusion}
Ben Poole, Ajay Jain, Jonathan~T Barron, and Ben Mildenhall.
\newblock Dreamfusion: Text-to-3d using 2d diffusion.
\newblock \emph{arXiv preprint arXiv:2209.14988}, 2022.

\bibitem[Raj et~al.(2018)Raj, Sangkloy, Chang, Hays, Ceylan, and Lu]{raj2018swapnet}
Amit Raj, Patsorn Sangkloy, Huiwen Chang, James Hays, Duygu Ceylan, and Jingwan Lu.
\newblock Swapnet: Image based garment transfer.
\newblock In \emph{Computer Vision--ECCV 2018: 15th European Conference, Munich, Germany, September 8--14, 2018, Proceedings, Part XII 15}, pages 679--695. Springer, 2018.

\bibitem[Ramesh et~al.(2021)Ramesh, Pavlov, Goh, Gray, Voss, Radford, Chen, and Sutskever]{ramesh2021zero}
Aditya Ramesh, Mikhail Pavlov, Gabriel Goh, Scott Gray, Chelsea Voss, Alec Radford, Mark Chen, and Ilya Sutskever.
\newblock Zero-shot text-to-image generation.
\newblock In \emph{International conference on machine learning}, pages 8821--8831. Pmlr, 2021.

\bibitem[Richardson et~al.(2023)Richardson, Metzer, Alaluf, Giryes, and Cohen-Or]{richardson2023texture}
Elad Richardson, Gal Metzer, Yuval Alaluf, Raja Giryes, and Daniel Cohen-Or.
\newblock Texture: Text-guided texturing of 3d shapes.
\newblock In \emph{ACM SIGGRAPH 2023 Conference Proceedings}, pages 1--11, 2023.

\bibitem[Rombach et~al.(2022)Rombach, Blattmann, Lorenz, Esser, and Ommer]{rombach2022high}
Robin Rombach, Andreas Blattmann, Dominik Lorenz, Patrick Esser, and Bj{\"o}rn Ommer.
\newblock High-resolution image synthesis with latent diffusion models.
\newblock In \emph{Proceedings of the IEEE/CVF conference on computer vision and pattern recognition}, pages 10684--10695, 2022.

\bibitem[Rong et~al.(2024)Rong, Grigorev, Wang, Black, Thomaszewski, Tsalicoglou, and Hilliges]{rong2024gaussian}
Boxiang Rong, Artur Grigorev, Wenbo Wang, Michael~J Black, Bernhard Thomaszewski, Christina Tsalicoglou, and Otmar Hilliges.
\newblock Gaussian garments: Reconstructing simulation-ready clothing with photorealistic appearance from multi-view video.
\newblock \emph{arXiv preprint arXiv:2409.08189}, 2024.

\bibitem[Saharia et~al.(2022)Saharia, Chan, Saxena, Li, Whang, Denton, Ghasemipour, Gontijo~Lopes, Karagol~Ayan, Salimans, et~al.]{saharia2022photorealistic}
Chitwan Saharia, William Chan, Saurabh Saxena, Lala Li, Jay Whang, Emily~L Denton, Kamyar Ghasemipour, Raphael Gontijo~Lopes, Burcu Karagol~Ayan, Tim Salimans, et~al.
\newblock Photorealistic text-to-image diffusion models with deep language understanding.
\newblock \emph{Advances in neural information processing systems}, 35:\penalty0 36479--36494, 2022.

\bibitem[Saito et~al.(2019)Saito, Huang, Natsume, Morishima, Kanazawa, and Li]{saito2019pifu}
Shunsuke Saito, Zeng Huang, Ryota Natsume, Shigeo Morishima, Angjoo Kanazawa, and Hao Li.
\newblock Pifu: Pixel-aligned implicit function for high-resolution clothed human digitization.
\newblock In \emph{Proceedings of the IEEE/CVF international conference on computer vision}, pages 2304--2314, 2019.

\bibitem[Sarafianos et~al.(2025)Sarafianos, Stuyck, Xiang, Li, Popovic, and Ranjan]{sarafianos2024garment3dgen}
Nikolaos Sarafianos, Tuur Stuyck, Xiaoyu Xiang, Yilei Li, Jovan Popovic, and Rakesh Ranjan.
\newblock Garment3dgen: 3d garment stylization and texture generation.
\newblock In \emph{3DV}, 2025.

\bibitem[Shi et~al.(2023{\natexlab{a}})Shi, Chen, Zhang, Liu, Xu, Wei, Chen, Zeng, and Su]{shi2023zero123++}
Ruoxi Shi, Hansheng Chen, Zhuoyang Zhang, Minghua Liu, Chao Xu, Xinyue Wei, Linghao Chen, Chong Zeng, and Hao Su.
\newblock Zero123++: a single image to consistent multi-view diffusion base model.
\newblock \emph{arXiv preprint arXiv:2310.15110}, 2023{\natexlab{a}}.

\bibitem[Shi et~al.(2023{\natexlab{b}})Shi, Wang, Ye, Long, Li, and Yang]{shi2023mvdream}
Yichun Shi, Peng Wang, Jianglong Ye, Mai Long, Kejie Li, and Xiao Yang.
\newblock Mvdream: Multi-view diffusion for 3d generation.
\newblock \emph{arXiv preprint arXiv:2308.16512}, 2023{\natexlab{b}}.

\bibitem[Song and Ermon(2019)]{song2019generative}
Yang Song and Stefano Ermon.
\newblock Generative modeling by estimating gradients of the data distribution.
\newblock \emph{Advances in neural information processing systems}, 32, 2019.

\bibitem[Tang et~al.(2023)Tang, Ren, Zhou, Liu, and Zeng]{tang2023dreamgaussian}
Jiaxiang Tang, Jiawei Ren, Hang Zhou, Ziwei Liu, and Gang Zeng.
\newblock Dreamgaussian: Generative gaussian splatting for efficient 3d content creation.
\newblock \emph{arXiv preprint arXiv:2309.16653}, 2023.

\bibitem[Tang et~al.(2025)Tang, Chen, Chen, Wang, Zeng, and Liu]{tang2025lgm}
Jiaxiang Tang, Zhaoxi Chen, Xiaokang Chen, Tengfei Wang, Gang Zeng, and Ziwei Liu.
\newblock Lgm: Large multi-view gaussian model for high-resolution 3d content creation.
\newblock In \emph{European Conference on Computer Vision}, pages 1--18. Springer, 2025.

\bibitem[Tung et~al.(2025)Tung, Chou, Cai, Yang, Zhang, Wetzstein, Hariharan, and Snavely]{tung2025megascenes}
Joseph Tung, Gene Chou, Ruojin Cai, Guandao Yang, Kai Zhang, Gordon Wetzstein, Bharath Hariharan, and Noah Snavely.
\newblock Megascenes: Scene-level view synthesis at scale.
\newblock In \emph{European Conference on Computer Vision}, pages 197--214. Springer, 2025.

\bibitem[Voleti et~al.(2025)Voleti, Yao, Boss, Letts, Pankratz, Tochilkin, Laforte, Rombach, and Jampani]{voleti2025sv3d}
Vikram Voleti, Chun-Han Yao, Mark Boss, Adam Letts, David Pankratz, Dmitry Tochilkin, Christian Laforte, Robin Rombach, and Varun Jampani.
\newblock Sv3d: Novel multi-view synthesis and 3d generation from a single image using latent video diffusion.
\newblock In \emph{European Conference on Computer Vision}, pages 439--457. Springer, 2025.

\bibitem[Wang et~al.(2023)Wang, Du, Li, Yeh, and Shakhnarovich]{wang2023score}
Haochen Wang, Xiaodan Du, Jiahao Li, Raymond~A Yeh, and Greg Shakhnarovich.
\newblock Score jacobian chaining: Lifting pretrained 2d diffusion models for 3d generation.
\newblock In \emph{Proceedings of the IEEE/CVF Conference on Computer Vision and Pattern Recognition}, pages 12619--12629, 2023.

\bibitem[Wang et~al.(2024)Wang, Sun, Tan, Chen, Chen, Li, Zhang, and Song]{wang2024towards}
Junyan Wang, Zhenhong Sun, Zhiyu Tan, Xuanbai Chen, Weihua Chen, Hao Li, Cheng Zhang, and Yang Song.
\newblock Towards effective usage of human-centric priors in diffusion models for text-based human image generation.
\newblock In \emph{Proceedings of the IEEE/CVF Conference on Computer Vision and Pattern Recognition (CVPR)}, pages 8446--8455, 2024.

\bibitem[Wang and Ye(2024)]{wang2024texfit}
Tongxin Wang and Mang Ye.
\newblock Texfit: Text-driven fashion image editing with diffusion models.
\newblock In \emph{Proceedings of the AAAI Conference on Artificial Intelligence}, pages 10198--10206, 2024.

\bibitem[Wang et~al.(2004)Wang, Bovik, Sheikh, and Simoncelli]{wang2004image}
Zhou Wang, Alan~C Bovik, Hamid~R Sheikh, and Eero~P Simoncelli.
\newblock Image quality assessment: from error visibility to structural similarity.
\newblock \emph{IEEE transactions on image processing}, 13\penalty0 (4):\penalty0 600--612, 2004.

\bibitem[Wang et~al.(2025)Wang, Wang, Chen, Xiang, Chen, Yu, Li, Su, and Zhu]{wang2025crm}
Zhengyi Wang, Yikai Wang, Yifei Chen, Chendong Xiang, Shuo Chen, Dajiang Yu, Chongxuan Li, Hang Su, and Jun Zhu.
\newblock Crm: Single image to 3d textured mesh with convolutional reconstruction model.
\newblock In \emph{European Conference on Computer Vision}, pages 57--74. Springer, 2025.

\bibitem[Watson et~al.(2022)Watson, Chan, Martin-Brualla, Ho, Tagliasacchi, and Norouzi]{watson2022novel}
Daniel Watson, William Chan, Ricardo Martin-Brualla, Jonathan Ho, Andrea Tagliasacchi, and Mohammad Norouzi.
\newblock Novel view synthesis with diffusion models.
\newblock \emph{arXiv preprint arXiv:2210.04628}, 2022.

\bibitem[Xiu et~al.(2022)Xiu, Yang, Tzionas, and Black]{xiu2022icon}
Yuliang Xiu, Jinlong Yang, Dimitrios Tzionas, and Michael~J Black.
\newblock Icon: Implicit clothed humans obtained from normals.
\newblock In \emph{2022 IEEE/CVF Conference on Computer Vision and Pattern Recognition (CVPR)}, pages 13286--13296. IEEE, 2022.

\bibitem[Xiu et~al.(2023)Xiu, Yang, Cao, Tzionas, and Black]{xiu2023econ}
Yuliang Xiu, Jinlong Yang, Xu Cao, Dimitrios Tzionas, and Michael~J Black.
\newblock Econ: Explicit clothed humans optimized via normal integration.
\newblock In \emph{Proceedings of the IEEE/CVF conference on computer vision and pattern recognition}, pages 512--523, 2023.

\bibitem[Xiu et~al.(2024)Xiu, Ye, Liu, Tzionas, and Black]{xiu2024puzzleavatar}
Yuliang Xiu, Yufei Ye, Zhen Liu, Dimitrios Tzionas, and Michael~J Black.
\newblock Puzzleavatar: Assembling 3d avatars from personal albums.
\newblock \emph{arXiv preprint arXiv:2405.14869}, 2024.

\bibitem[Xu et~al.(2024{\natexlab{a}})Xu, Cheng, Gao, Wang, Gao, and Shan]{xu2024instantmesh}
Jiale Xu, Weihao Cheng, Yiming Gao, Xintao Wang, Shenghua Gao, and Ying Shan.
\newblock Instantmesh: Efficient 3d mesh generation from a single image with sparse-view large reconstruction models.
\newblock \emph{arXiv preprint arXiv:2404.07191}, 2024{\natexlab{a}}.

\bibitem[Xu et~al.(2024{\natexlab{b}})Xu, Shi, Yifan, Chen, Yang, Peng, Shen, and Wetzstein]{xu2024grm}
Yinghao Xu, Zifan Shi, Wang Yifan, Hansheng Chen, Ceyuan Yang, Sida Peng, Yujun Shen, and Gordon Wetzstein.
\newblock Grm: Large gaussian reconstruction model for efficient 3d reconstruction and generation.
\newblock \emph{arXiv preprint arXiv:2403.14621}, 2024{\natexlab{b}}.

\bibitem[Yang et~al.(2016)Yang, Ambert, Pan, Wang, Yu, Berg, and Lin]{yang2016detailed}
Shan Yang, Tanya Ambert, Zherong Pan, Ke Wang, Licheng Yu, Tamara Berg, and Ming~C Lin.
\newblock Detailed garment recovery from a single-view image.
\newblock \emph{arXiv preprint arXiv:1608.01250}, 2016.

\bibitem[Yang et~al.(2024{\natexlab{a}})Yang, Shi, Zhang, Yang, Wang, Zhao, Liu, Wang, Lin, Yu, et~al.]{yang2024hunyuan3d}
Xianghui Yang, Huiwen Shi, Bowen Zhang, Fan Yang, Jiacheng Wang, Hongxu Zhao, Xinhai Liu, Xinzhou Wang, Qingxiang Lin, Jiaao Yu, et~al.
\newblock Hunyuan3d-1.0: A unified framework for text-to-3d and image-to-3d generation.
\newblock \emph{arXiv preprint arXiv:2411.02293}, 2024{\natexlab{a}}.

\bibitem[Yang et~al.(2024{\natexlab{b}})Yang, Shi, Zhang, Yang, Wang, Zhao, Liu, Wang, Lin, Yu, et~al.]{yang2024tencent}
Xianghui Yang, Huiwen Shi, Bowen Zhang, Fan Yang, Jiacheng Wang, Hongxu Zhao, Xinhai Liu, Xinzhou Wang, Qingxiang Lin, Jiaao Yu, et~al.
\newblock Tencent hunyuan3d-1.0: A unified framework for text-to-3d and image-to-3d generation.
\newblock \emph{arXiv e-prints}, pages arXiv--2411, 2024{\natexlab{b}}.

\bibitem[Yu et~al.(2024)Yu, Duan, Hur, Sargent, Rubinstein, Freeman, Cole, Sun, Snavely, Wu, et~al.]{yu2024wonderjourney}
Hong-Xing Yu, Haoyi Duan, Junhwa Hur, Kyle Sargent, Michael Rubinstein, William~T Freeman, Forrester Cole, Deqing Sun, Noah Snavely, Jiajun Wu, et~al.
\newblock Wonderjourney: Going from anywhere to everywhere.
\newblock In \emph{Proceedings of the IEEE/CVF Conference on Computer Vision and Pattern Recognition}, pages 6658--6667, 2024.

\bibitem[Zhang et~al.(2024)Zhang, Wang, Carrasco, Wu, Yang, Beeler, and la~Torre]{zhang2024sa}
Cheng Zhang, Yuanhao Wang, Francisco~Vicente Carrasco, Chenglei Wu, Jinlong Yang, Thabo Beeler, and Fernando~De la Torre.
\newblock {FabricDiffusion}: High-fidelity texture transfer for 3d garments generation from in-the-wild images.
\newblock In \emph{ACM SIGGRAPH Asia}, 2024.

\bibitem[Zhang et~al.(2018)Zhang, Isola, Efros, Shechtman, and Wang]{zhang2018unreasonable}
Richard Zhang, Phillip Isola, Alexei~A Efros, Eli Shechtman, and Oliver Wang.
\newblock The unreasonable effectiveness of deep features as a perceptual metric.
\newblock In \emph{Proceedings of the IEEE conference on computer vision and pattern recognition}, pages 586--595, 2018.

\bibitem[Zheng et~al.(2021)Zheng, Yu, Liu, and Dai]{zheng2021pamir}
Zerong Zheng, Tao Yu, Yebin Liu, and Qionghai Dai.
\newblock Pamir: Parametric model-conditioned implicit representation for image-based human reconstruction.
\newblock \emph{IEEE transactions on pattern analysis and machine intelligence}, 44\penalty0 (6):\penalty0 3170--3184, 2021.

\bibitem[Zhou et~al.(2013)Zhou, Chen, Fu, Guo, and Tan]{zhou2013garment}
Bin Zhou, Xiaowu Chen, Qiang Fu, Kan Guo, and Ping Tan.
\newblock Garment modeling from a single image.
\newblock In \emph{Computer graphics forum}, pages 85--91. Wiley Online Library, 2013.

\bibitem[Zhou et~al.(2010)Zhou, Fu, Liu, Cohen-Or, and Han]{zhou2010parametric}
Shizhe Zhou, Hongbo Fu, Ligang Liu, Daniel Cohen-Or, and Xiaoguang Han.
\newblock Parametric reshaping of human bodies in images.
\newblock \emph{ACM transactions on graphics (TOG)}, 29\penalty0 (4):\penalty0 1--10, 2010.

\bibitem[Zhou and Tulsiani(2023)]{zhou2023sparsefusion}
Zhizhuo Zhou and Shubham Tulsiani.
\newblock Sparsefusion: Distilling view-conditioned diffusion for 3d reconstruction.
\newblock In \emph{Proceedings of the IEEE/CVF Conference on Computer Vision and Pattern Recognition}, pages 12588--12597, 2023.

\bibitem[Zhu et~al.(2022)Zhu, Qiu, Qiu, and Han]{zhu2022registering}
Heming Zhu, Lingteng Qiu, Yuda Qiu, and Xiaoguang Han.
\newblock Registering explicit to implicit: Towards high-fidelity garment mesh reconstruction from single images.
\newblock In \emph{Proceedings of the IEEE/CVF Conference on Computer Vision and Pattern Recognition}, pages 3845--3854, 2022.

\bibitem[Zhu et~al.(2024)Zhu, Li, Liu, Peng, Yang, and Kemelmacher-Shlizerman]{zhu2024m}
Luyang Zhu, Yingwei Li, Nan Liu, Hao Peng, Dawei Yang, and Ira Kemelmacher-Shlizerman.
\newblock M\&m vto: Multi-garment virtual try-on and editing.
\newblock In \emph{Proceedings of the IEEE/CVF Conference on Computer Vision and Pattern Recognition}, pages 1346--1356, 2024.

\bibitem[Zou et~al.(2023)Zou, Han, and Wong]{zou2023cloth4d}
Xingxing Zou, Xintong Han, and Waikeung Wong.
\newblock Cloth4d: A dataset for clothed human reconstruction.
\newblock In \emph{Proceedings of the IEEE/CVF Conference on Computer Vision and Pattern Recognition}, pages 12847--12857, 2023.

\end{thebibliography}
}
\clearpage

\maketitlesupplementary
\appendix
\renewcommand\thesection{\Alph{section}}
\renewcommand{\thetable}{S\arabic{table}}  
\renewcommand{\thefigure}{S\arabic{figure}}



\section{Contribution, Novelty, and Limitation}
We reiterate our contribution, novelty, and limitations.

\mypara{Contribution and Novelty.}
Our main contribution lies in a new direction: enabling non-professional users to create and edit 3D garment with single-view input. While existing works have made strides in reconstructing clothed humans~\cite{saito2019pifu,zheng2021pamir,alldieck2022photorealistic,xiu2023econ} or garment~\cite{sarafianos2024garment3dgen} from a single image, they mainly rely on optimizing pre-defined garment or human templates. In contrast, we target a more flexible, template-free garment reconstruction framework. Specifically, we propose to progressively synthesize depth-accurate novel view images with enhanced cross-view consistency. Moreover, our method enables single-view 3D editing, including part-based or local surface edits --- capabilities that are absent in the aforementioned methods.

\mypara{Scope and Limitations.}
As discussed in Section 5 of the main paper, our method has certain limitations. We mainly focus on garment in a rest pose. As will be shown in Section~\ref{sec:suppl_failure}, our method may struggle to accurately capture the geometry of garments in non-rest poses. With that said, this scope is a deliberate choice, as rest poses provide a consistent and intuitive baseline that aligns well with the needs of garment editing applications.

\section{Ethics and Social Impacts}
We focus on advancing garment digitization. We do not foresee any ethical concerns or negative societal impacts arising from our work. Our training and evaluation processes do not involve any sensitive data, human identities, or personal information. All experiments and datasets used in this study are compliant with ethical research practices. By advancing template-free garment reconstruction for non-professional users, our method avoids potential biases associated with specific body or garment templates, promoting inclusiveness in digital garment reconstruction.

\section{Additional Implementation Details}
\label{sec:appendixA}

In this section, we provide additional implementation details of our method omitted in the main text.

\subsection{Conditional Image Generation}

Our image generation model is finetuned from the Stable Zero-1-to-3 checkpoint\footnote{https://huggingface.co/stabilityai/stable-zero123}. To account for the additional projected image as input, we add 4 additional channels to the input convolution layer of the denosing UNet and initialize the weights to be zeros. The training resolution is 512$\times$512. We train the mode on 4 NVIDIA A6000 GPUs with a total batch size of 256 for 20k iterations for 2 days.

\subsection{Conditional Depth Generation}

Our conditional image generation model is finetuned from the Sapiens-0.3B depth checkpoint\footnote{https://huggingface.co/facebook/sapiens-depth-0.3b}. To add the projected partial depth map as the additional condition, we add 1 extra channels to the input projection layer of the vision transformer backbone and initialize its weights to be zeros. The training resolution is 512$\times$512. We train the model on 4 A6000 GPUs with a total batch size of 24 for 3 days.

\subsection{Computational Efficiency}
The inference time and memory consumption of our method are approximately 1 minute and 10 GB, respectively, on a single A6000 GPU. These values are comparable to those of most baseline methods, which have inference times ranging from 10 seconds to 1 minute.

\subsection{Measures to Reduce Error Accumulation}

Since our method synthesizes novel views in sequential steps, it is susceptible to error accumulation. To address this, we incorporate a series of techniques aimed at mitigating such errors and improving overall robustness.

\mypara{Point Cloud Outlier Removal.}
Depth predictions near the edges of discontinuities (with large jumps in depth values) are occasionally inaccurate, resulting in some floating points in the point cloud. To address this, we apply a classical outlier removal method at each step to eliminate these floating points, ensuring a cleaner and accurate point cloud.

\begin{figure}[t]
    \centering
    \vspace{1mm}
    \includegraphics[width=1\linewidth]{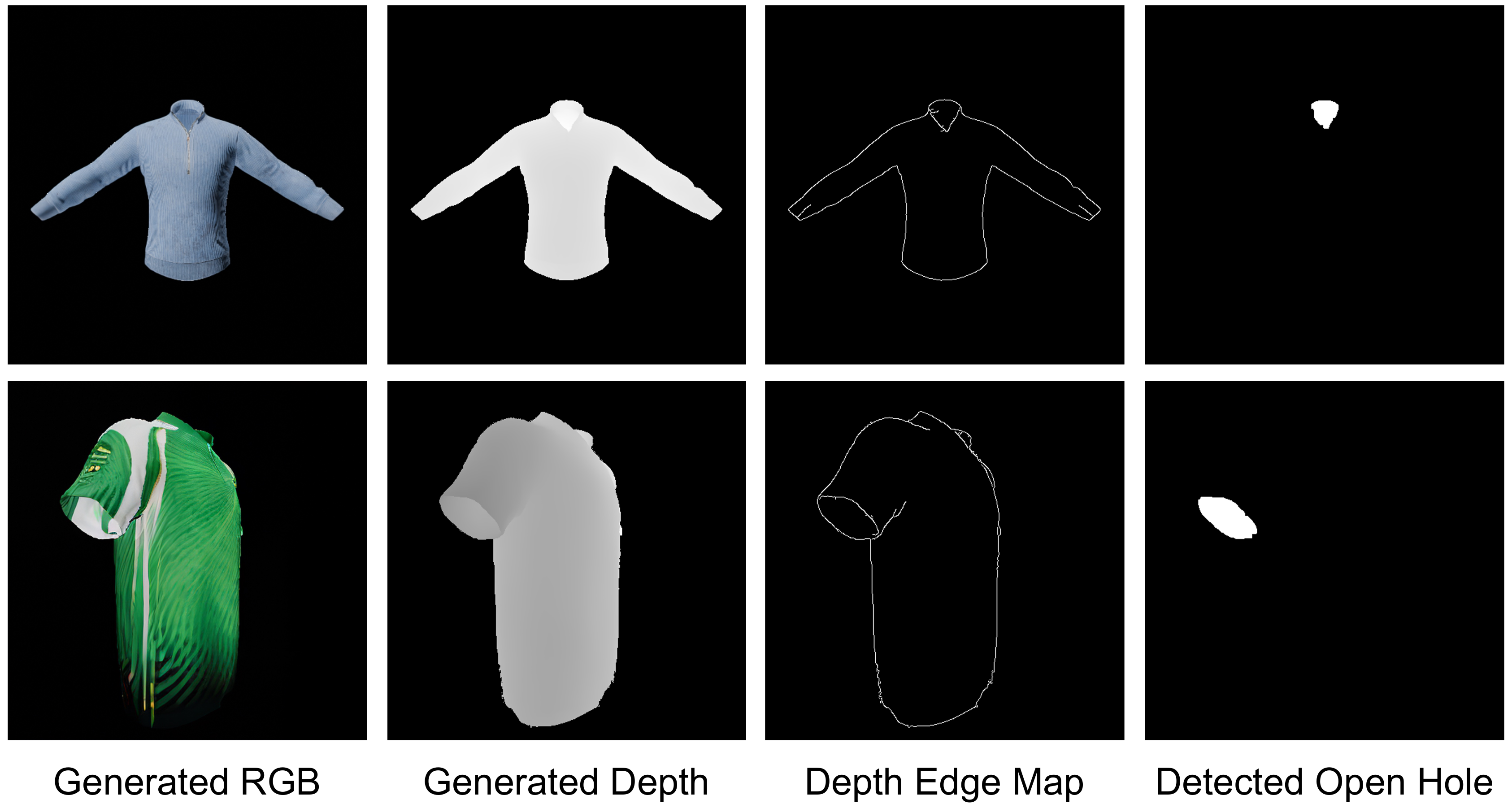}
    \vspace{-7mm}
    \caption{\small \textbf{Open hole detection in garments.} We note that interior regions of open holes in a garment exhibit greater depth values compared to the boundary pixels. Leveraging this observation, we propose a simple yet effective algorithm to detect open holes and exclude these regions during point cloud completion, improving the robustness of the pipeline.
    } 
    \label{fig:suppl_hole}
\end{figure}

\mypara{Open Hole Detection.}
We observe that depth predictions are less reliable in open-hole regions of a garment surface, such as holes in collars and sleeves. Additionally, the surface orientation derived from the estimated depth map in these areas can be reversed. These errors can propagate and lead to artifacts in subsequent steps. To address this issue, we develop a simple algorithm to detect open holes and exclude these regions during point cloud completion, improving the robustness of the pipeline.

The detection algorithm is based on the observation that the interior regions of open holes typically exhibit greater depth values compared to the boundary pixels. As shown in Figure~\ref{fig:suppl_hole}, after synthesizing the completed image and depth maps from a novel viewpoint, we first detect edges in the depth map and identify connected regions enclosed by these edges using classical methods. A connected region $R$ is classified as an open hole if more than a threshold $\epsilon$ of its boundary pixels have depth values smaller than the average depth of the region. For all our experiments, we found that $\epsilon$ can be robustly set to 0.85.

\mypara{Clipping Distant Depth Values.}
Our observations indicate that synthesized images and depth maps are more robust in regions closer to the camera compared to those farther away. At steps 3 and 4 (corresponding to azimuth angles of $120^\circ$ and $-120^\circ$), the entire back side of the garment is synthesized from a side view. For these steps, we only use pixels with smaller depth values for point cloud completion, disregarding pixels with larger depth values.

\subsection{Point-to-Mesh Reconstruction}
We use Screened Poisson surface reconstruction to convert point clouds to meshes. Note that the point orientations are estimated using depth maps at each step, as described in Section 3.3 of the main paper. While Screened Poisson reconstruction generates a watertight mesh, we aim to preserve the non-watertight topology of garments (e.g., maintaining holes in collars, sleeves, etc.). To achieve this, we perform an additional trimming operation\footnote{https://github.com/mkazhdan/PoissonRecon} to remove unwanted mesh faces introduced during the Poisson reconstruction that fill open holes by leveraging the point cloud density.
To further reduce artifacts, we remove floating faces unconnected to the main mesh and apply Laplacian smoothing to refine the mesh surface.

\begin{figure*}[t]
    \centering
    \includegraphics[width=1\linewidth]{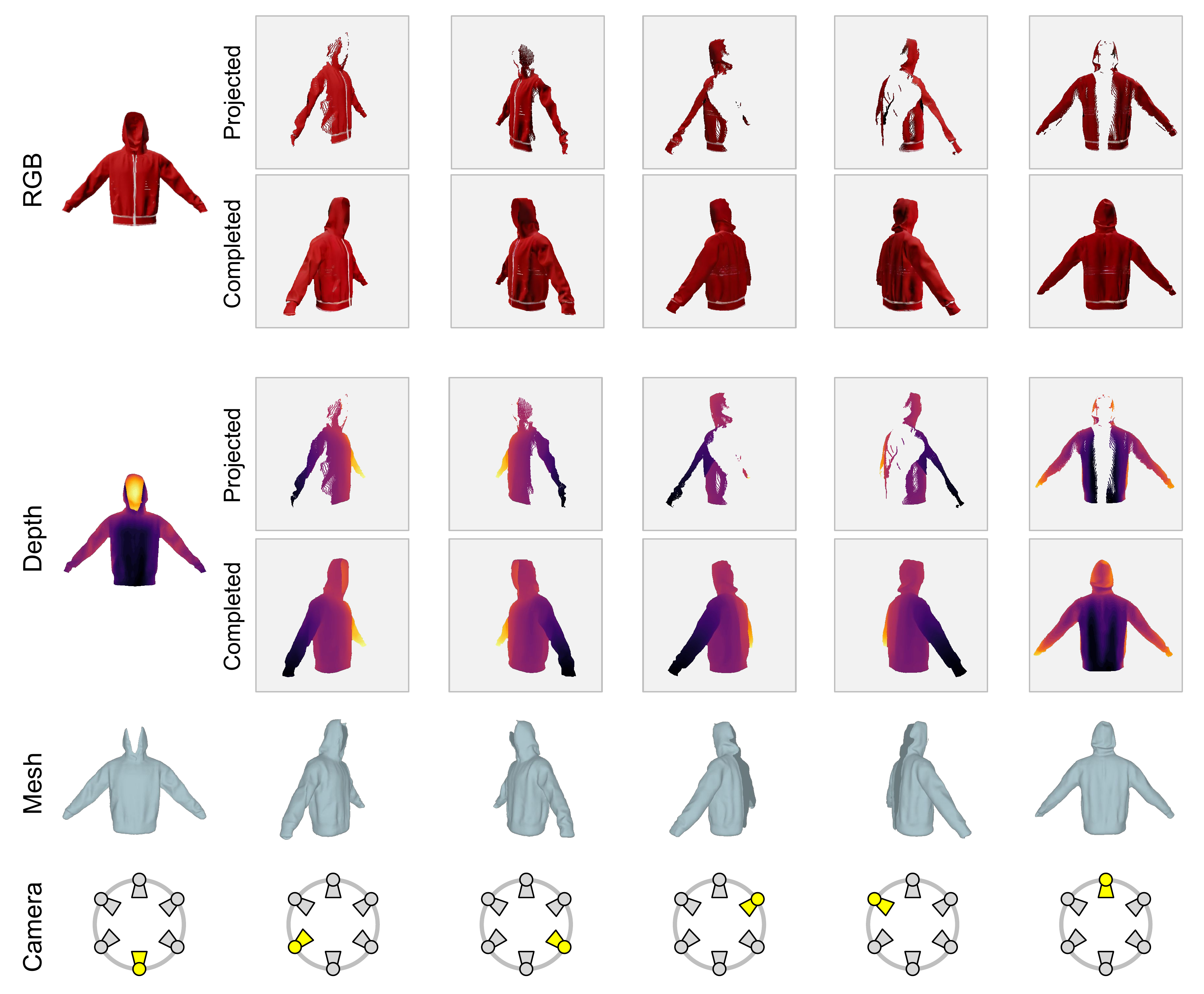}
    \vspace{-6.6mm}
    \caption{\small \textbf{Intermediate results of progressive novel view synthesis along a full camera trajectory.} From an input RGB image (top-left), \ourmethod progressively synthesize novel view RGB and depth maps following a zigzag camera trajectory.
    } 
    \vspace{-3mm}
    \label{fig:suppl_full_results}
\end{figure*}

\subsection{Scope of Single-View 3D Editing}
As introduced in Section 3.3 of the main paper, \ourmethod enables single-view editing through a simple workflow: identify the edited 3D region, remove the original mesh in the identified area, and reconstruct the edited components. We support two types of editing operations, differentiated by their assumptions about the edited regions.

The first category is local surface editing. Given a camera viewpoint and a mask, this approach assumes that only the visible surface intersected by the camera rays corresponding to the masked pixels will be edited. Occluded surfaces are ignored, even if their mesh vertices project within the mask. To facilitate reconstruction, we remove the mesh vertices of the selected surface. Additionally, internal vertices near the external surface are also removed to account for surface thickness.

The second category, part-based editing, involves modifying a 3D garment part, including not only the ``front'' surface but also the ``back'' and ``internal" surfaces within a masked region. For ease of implementation, we always use the frontal view as the editing perspective and remove all mesh vertices whose 2D projections fall within the mask.

Our editing pipeline is designed under the assumption that both the geometry and the texture will be edited. Therefore, it is not optimized for cases where (1) surface texture is modified while preserving the geometry, or (2) the geometry or pose is altered while preserving the texture.

\subsection{2D Editing Assumptions}
In theory, our method is agnostic to the tools used for 2D editing. The edits can be created using deep learning-based image editing models or traditional tools like Photoshop. However, our approach requires the edits to be confined to regions specified by masks in the 2D input. Therefore, global edits such as style transfer that alters the entire image, are not recommended.

\section{Additional Results and Analyses}
\subsection{Intermediate Results of Progressive NVS}
In Figure 2 of the main paper, we showed results at one specific camera rotation step during the progressive novel view synthesis. Here, we illustrate the whole process and show the intermediate results in Figure~\ref{fig:suppl_full_results}.

\subsection{Additional Baseline Comparisons}
\label{sec:appendixD2}

\subsubsection{Comparison with SoTA NVS methods}
We present additional quantitative comparisons for novel view synthesis against state-of-the-art methods (Zero-1-to-3++~\cite{shi2023zero123++} \& MVD-Fusion~\cite{hu2024mvd}, fine-tuned with same data).
For each object in the held-out test set of 150 garment assets, we sample six camera viewpoints with an elevation of 20 degrees and evenly spaced azimuth angles covering 360 degrees. Each method takes a frontal image as input and generates six corresponding novel views, which we evaluate against ground truth images using image similarity metrics (LPIPS, PSNR, and SSIM). We also report our proposed CVCS score. Table~\ref{tab:nvs} shows that our method achieves superior performance across all metrics.

\subsubsection{Qualitative Comparison with Garment3DGen}
\begin{figure}[t]
    \centering
    \includegraphics[width=8cm]{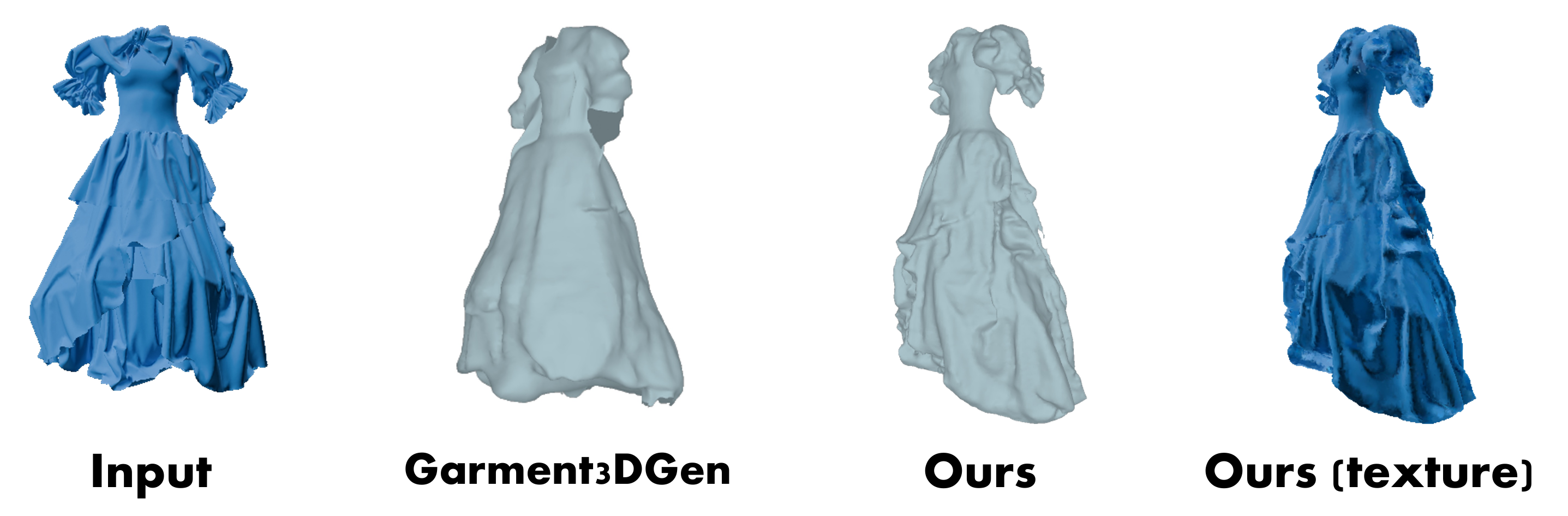}
    \vspace{-3mm}
    \caption{\textbf{Qualitative comparison with Garment3DGen}~\cite{sarafianos2024garment3dgen}. Our \ourmethod reconstructs garment meshes with richer details with much lower computational costs.}
    \label{fig:rebuttal}
     \vspace{-1mm}
\end{figure}

As the texture reconstruction code of Garment3DGen~\cite{sarafianos2024garment3dgen} is not released, we provide qualitative comparison with Garment3DGen on the reconstructed mesh geometry in Figure~\ref{fig:rebuttal}. Our method reconstructs 3D garments with much richer geometric details and much less inference time (1 min vs. 3 hours).

\begin{table}[t]
\small
\tabcolsep 6.5pt 
\centering
\caption{\textbf{Quantitative comparison for novel view synthesis.} Our method outperforms all state-of-the-art novel view synthesis methods cross both image similarity and consistency metrics.}
\vspace{-2mm}
\begin{tabular}{c|ccc|c}
\toprule
 & LPIPS $\downarrow$ & PSNR $\uparrow$ & SSIM $\uparrow$ & CVCS$\uparrow$ \\
\midrule
Zero123++ & 0.1611 & 18.023 & 0.7979 & 0.8957 \\
MVD-Fusion & 0.1528 & 18.529 & 0.8026 & 0.9090 \\
Ours & \textbf{0.1052} & \textbf{22.776} & \textbf{0.8557} & \textbf{0.9512} \\
\bottomrule
\end{tabular} \label{tab:nvs}
\vspace{-2mm}
\end{table}

\subsection{Additional Analyses and Applications}
\label{sec:appendixD}

\subsubsection{Degree of Zigzag Camera Trajectory}
We have studied all major design choices in our pipeline in the main paper, including the effect of progressive novel view synthesis and camera trajectory. Here, we analyze the impact of the degree of Zigzag Camera Trajectory and show the results in Table~\ref{tab:suppl_ablation_degree}. In our experiments, we use a 60$^{\circ}$ trajectory as it provides a good balance between view coverage and efficiency. While the choice of degree slightly affects the ability to synthesize side-view garments (i.e., 90$^{\circ}$), our analysis indicates that the overall performance is not highly sensitive to this parameter. We do not notice any other significant hyperparameters in our framework.

\begin{table}[t]
\small
\tabcolsep 9pt 
\centering
\caption{\small \textbf{Analysis of the degree of zigzag camera trajectory.} In our experiments, we use a 60$^{\circ}$ trajectory as it provides a good balance between view coverage and efficiency. While the choice of degree slightly affects the ability to synthesize side-view garments (i.e., 90$^{\circ}$), our analysis indicates that the overall performance is not highly sensitive to this parameter.}
\begin{tabular}{c|ccc|c}
\toprule
\multirow{2.5}{*}{Degree} & \multicolumn{3}{c|}{Appearance} & \multicolumn{1}{l}{Geometry} \\
\cmidrule{2-5}
 & SSIM$\uparrow$ & LPIPS$\downarrow$ & PSNR $\uparrow$ & Chamfer$\downarrow$  \\
\midrule
30$^{\circ}$ & 0.8044 & 0.1675 & \textbf{20.62} & 0.0051 \\
60$^{\circ}$ & \textbf{0.8066} & \textbf{0.1638} & \underline{20.62} & \textbf{0.0050} \\
90$^{\circ}$ & 0.8003 & 0.1709 & 20.19 & 0.0070 \\
120$^{\circ}$ & \underline{0.8053} & \underline{0.1654} & 20.51 & \underline{0.0050} \\
\bottomrule
\end{tabular}
\label{tab:suppl_ablation_degree}
\end{table}

\begin{figure}[t]
    \centering
    \vspace{-5mm}
    \includegraphics[width=0.95\linewidth]{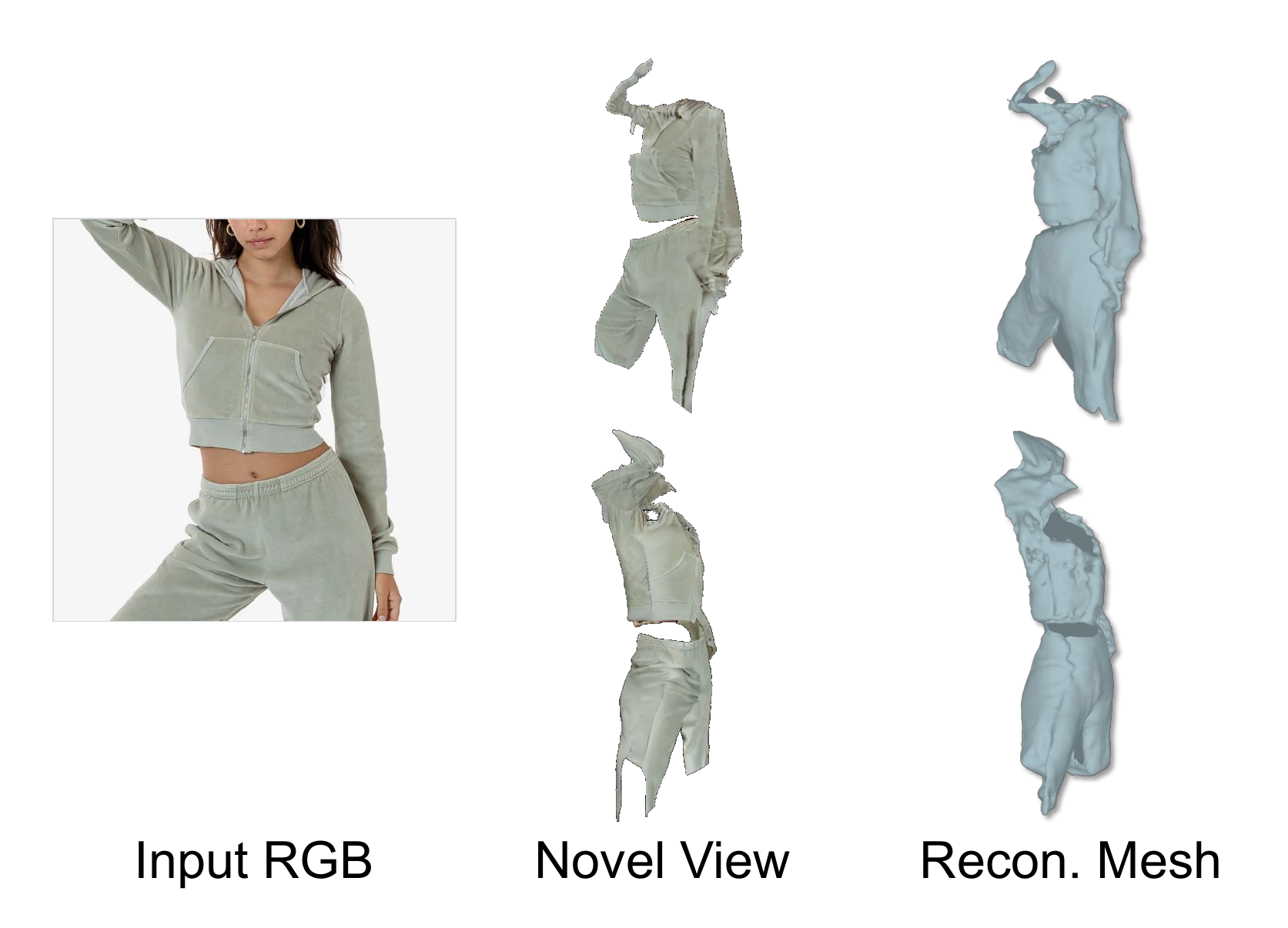}
    \vspace{-3mm}
    \caption{\small \textbf{Failure case.} \ourmethod may fail to reconstruct the garment with arbitrary poses. 
    } 
    \vspace{-3mm}
    \label{fig:supplement_limitation}
\end{figure}

\begin{figure*}[t]
    \centering
    \includegraphics[width=0.93\linewidth]{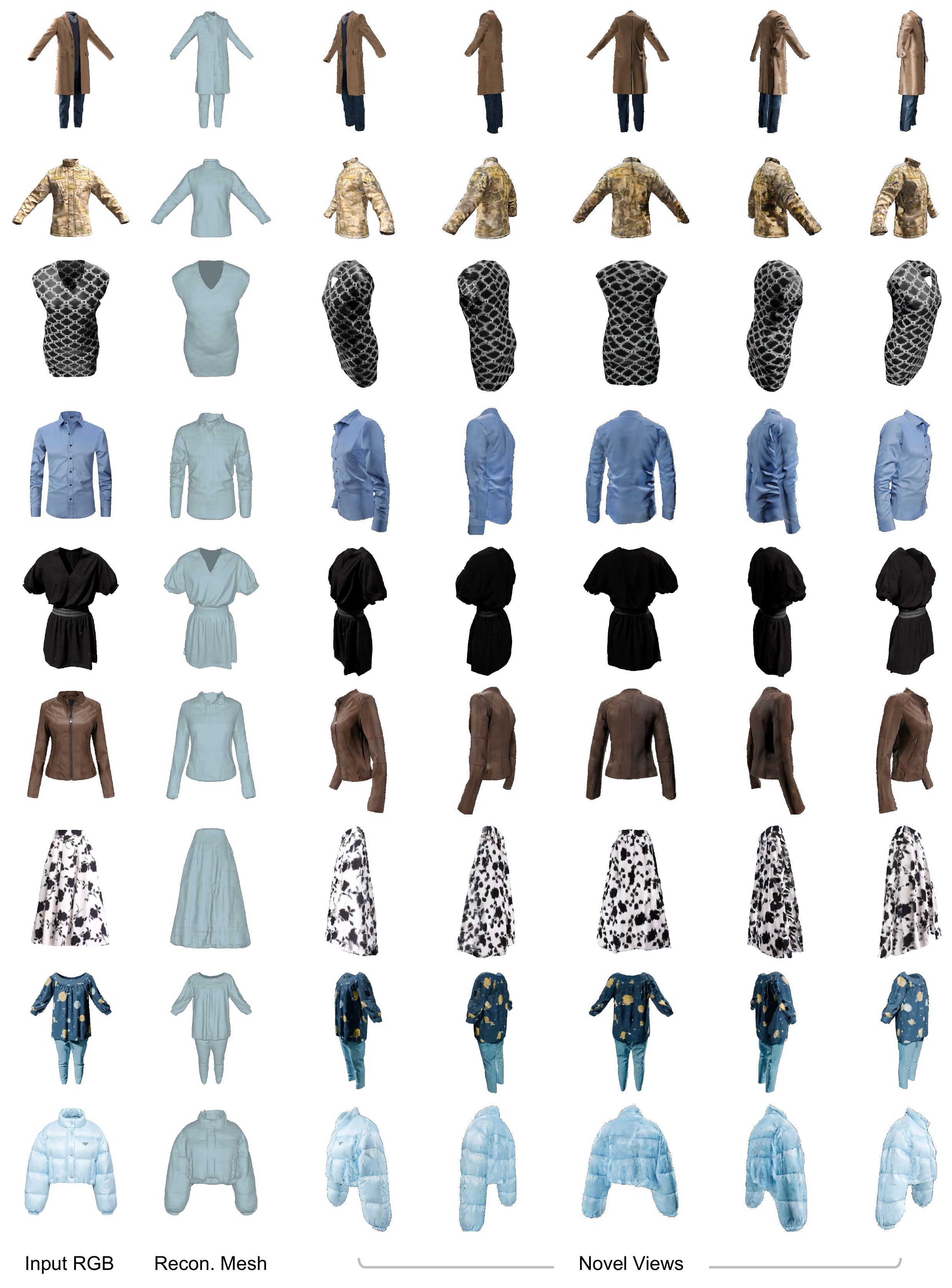}
    \caption{\small More qualitative result on single-view 3D garment reconstruction.
    } 
    \label{fig:suppl_more_recon}
\end{figure*}

\begin{figure*}[t]
    \centering
    \includegraphics[width=1\linewidth]{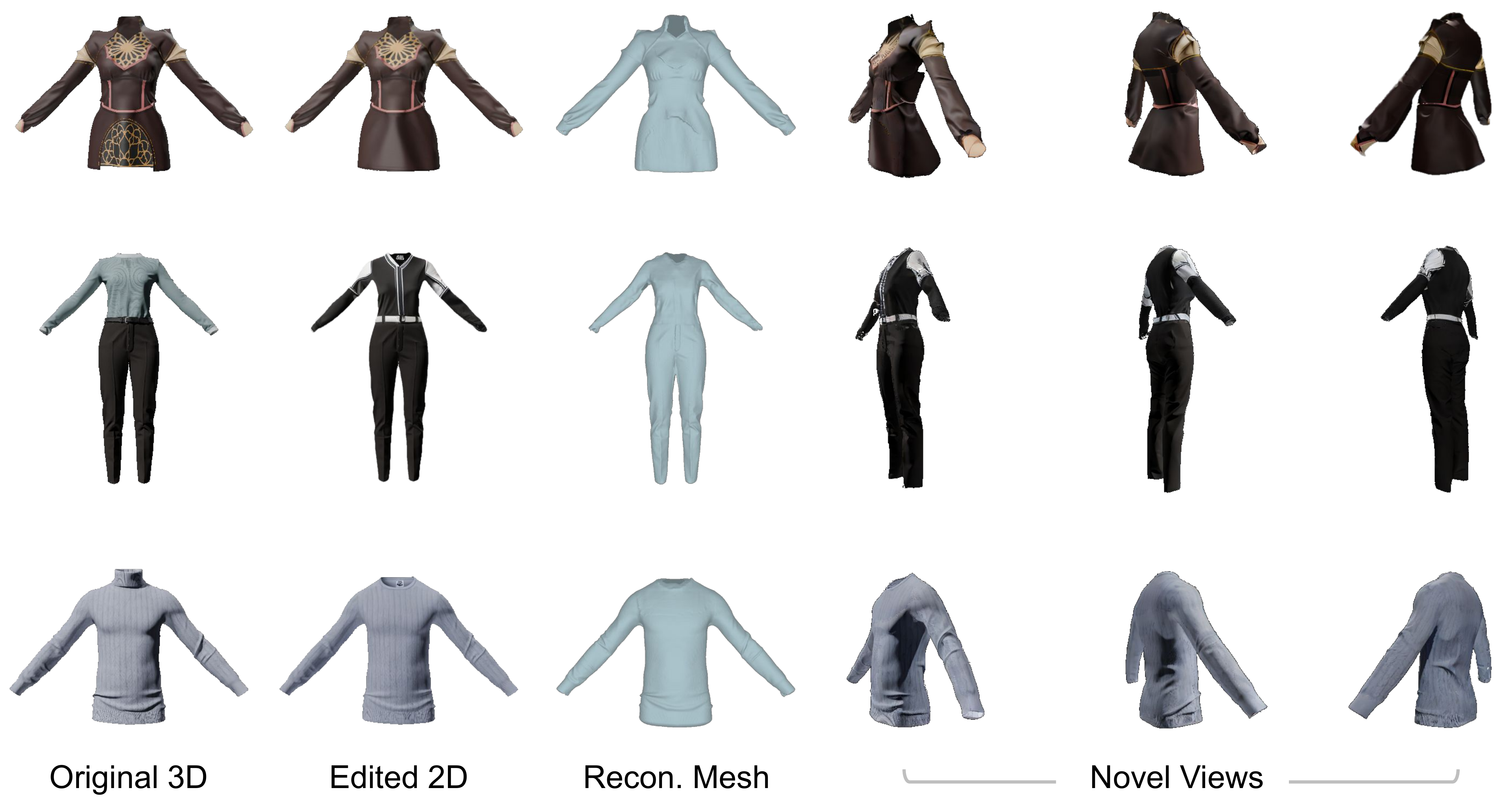}
    \caption{\small \textbf{More results on single-view 3D garment editing.} The top row illustrates how \ourmethod effectively handles surface edits, even for regions with complex textures. The middle row demonstrates the capability of \ourmethod to support full garment changes and swaps, showcasing the potential in virtual try-on scenarios. The bottom row presents an example of removing an entire garment part.
    } 
    \vspace{5mm}
    \label{fig:supplement_more_edit}
\end{figure*}

\begin{figure*}[t]
    \centering
    \includegraphics[width=1\linewidth]{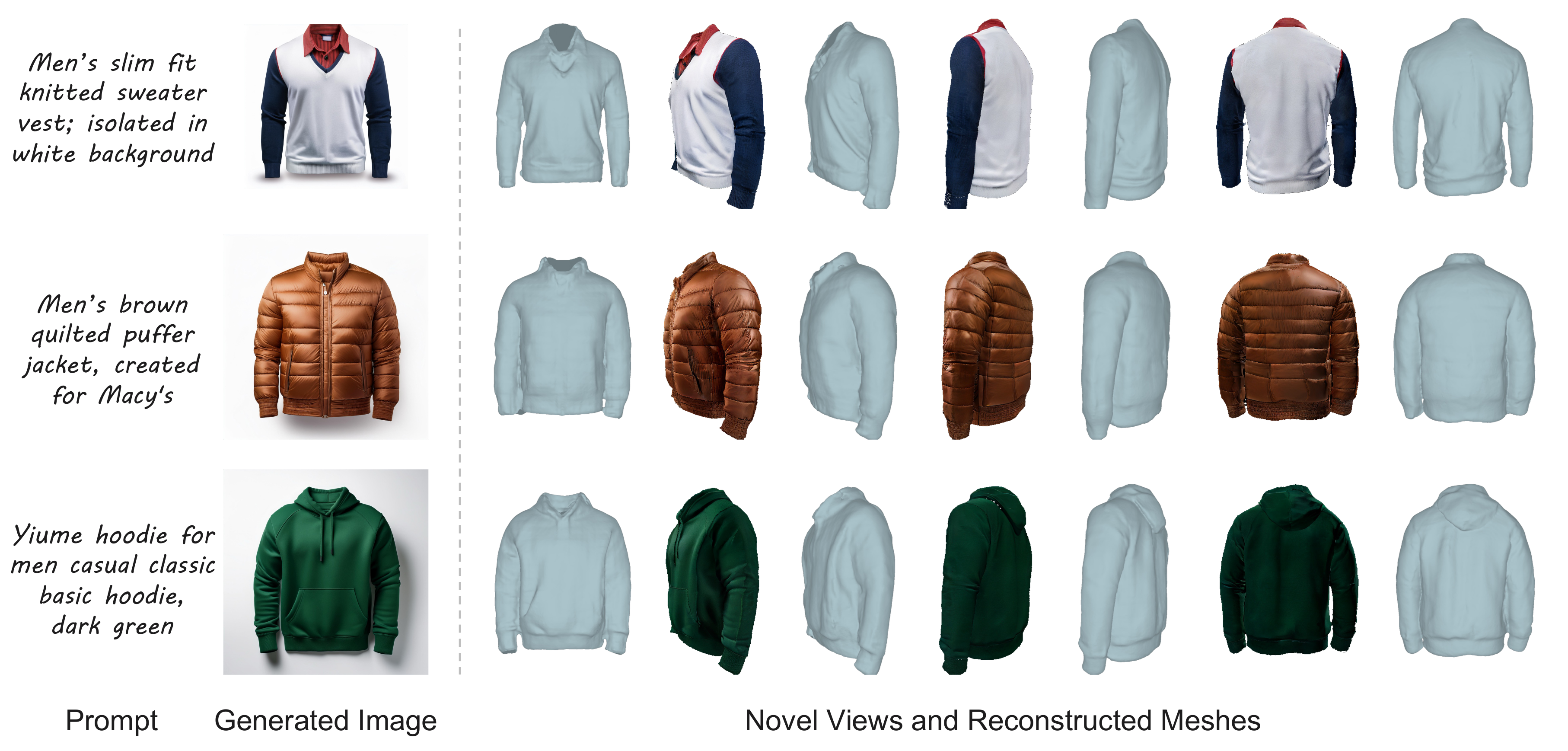}
    \caption{\small \textbf{Compatibility with generative apparel.} By reconstructing both geometry and texture from synthetic garment images, \ourmethod demonstrates its adaptability to AI-generated designs. The results showcase the ability of \ourmethod to handle diverse and complex inputs, expanding its potential applications to generative fashion and virtual apparel workflows.
    } 
    \label{fig:supplement_ai_gen}
\end{figure*}

\subsubsection{Digitizing AI-generated Apparel}
We explore the potential of combing \ourmethod with AI-generated garment image and show examples in Figure~\ref{fig:supplement_ai_gen}. Using a text-to-image generative model, we produce synthetic garment images and apply \ourmethod to digitize them. The results demonstrate the broad applicability of our method in handling diverse inputs, including AI-generated designs.

\subsection{Failure Cases}\label{sec:suppl_failure}
The focus of our work is on reconstructing and editing garments in their rest pose. Consequently, our method struggles with input images in arbitrary poses as such instances lie outside of the training data distribution. As illustrated in Figure~\ref{fig:supplement_limitation}, an input garment image in a non-resting pose results in the failure of our model to synthesize coherent novel view images, leading to nonsensical reconstructions.

\subsection{More Qualitative Results}
\label{sec:appendixC}

\mypara{Reconstruction.}
Please see more results in Figure~\ref{fig:suppl_more_recon}.

\mypara{Editing.}
We provide more qualitative results in Figure~\ref{fig:supplement_more_edit}.

\end{document}